\documentclass{article}
\usepackage{amsmath,amssymb}
\usepackage{mathdots}
\usepackage{mathrsfs}
\usepackage{bm}
\usepackage{epsfig}
\usepackage[centerlast]{subfigure}
\usepackage{graphicx}
\usepackage{amsmath}
\usepackage{psfig}

\usepackage{ntheorem}

\def\nsection#1{\section{#1}\setcounter{equation}{0}}

\renewcommand{\theequation}{\thesection.\arabic{equation}}
\newcommand{\eq}[1]{\begin{equation} #1 \end{equation}} 

\newcommand{\Ai}{{\mathrm{Ai}}}
\newcommand{\Bi}{{\mathrm{Bi}}}  
\newcommand{\StNo}{{\mathit{St}}}

\newcommand{\qq}{\begin{eqnarray}}
\newcommand{\qqq}{\end{eqnarray}}
\newcommand{\ee}{{\rm e}}

\newcommand{\CV}{{\cal V}}

\newcommand{\Id}{{I\hspace{-0.03cm}d}}
\newcommand{\m}{\hspace{0.025cm}}

\newcommand{\dd}{{\mathrm{d}}} 
\newcommand{\RR}{{\mathbb{R}}} 
\newcommand{\const}{{\mathrm{const.}}} 
\newcommand{\ReNo}{{\mathit{Re}}} 
\newcommand{\half}{{\textstyle\frac12}} 
\newcommand{\third}{{\ensuremath{3^\text{rd}}} } 
\newcommand{\eff}{{\text{eff}}} 
\newcommand{\scprod}{\cdot}

\newcommand{\antivec}[1]{\ensuremath\reflectbox{\hspace{-0pt}$\vec{\text{\reflectbox{$\!#1\!\,$}}}\,$}}

\theoremstyle{plain}

\newtheorem*{conc}{{Conclusion.}} 
\theorembodyfont{\rm} 
\theoremstyle{break}

\begin{document}

\title{\textbf{Stochastic processes in turbulent 
transport}\thanks{\hspace*{0.1cm}Text of lectures at the School 
on Stochastic Geometry, the Stochastic Loewner Evolution, and 
Non-Equilibrium\hfill\break\hspace*{0.7cm}Growth Processes, Trieste, Italy,
July 7-18, 2008}}
\author{Krzysztof Gaw\c{e}dzki 
\\
\small{Laboratoire de Physique, C.N.R.S., ENS-Lyon,
Universit\'e de Lyon,}\cr
\small{ 46 All\'ee d'Italie, 69364 Lyon, France}}
\date{}
\maketitle

\abstract{\noindent This is a set of four lectures devoted 
to simple ideas about turbulent transport, a ubiquitous non-equilibrium 
phenomenon. In the course similar to that given by the author in 2006 
in Warwick \cite{Warwick}, we discuss lessons which have been learned 
from naive models of turbulent mixing that employ simple random velocity 
ensembles and study related stochastic processes. In the first lecture, 
after a brief reminder of the turbulence phenomenology, we describe 
the intimate relation between the passive advection of particles and 
fields by hydrodynamical flows. The second lecture is devoted to some 
useful tools of  the multiplicative ergodic theory for random dynamical 
systems. In the third lecture, we apply these tools to the example 
of particle flows in the Kraichnan ensemble of smooth velocities that 
mimics turbulence at intermediate Reynolds numbers. In the fourth lecture, 
we extends the discussion of particle flows to the case of non-smooth 
Kraichnan velocities that model fully developed turbulence. We stress 
the unconventional aspects of particle flows that appear in this regime 
and lead to phase transitions in the presence of compressibility. 
The intermittency of scalar fields advected by fully turbulent velocities 
and the scenario linking it to hidden statistical conservation laws 
of multi-particle flows are briefly explained.}

\ 

\

\nsection{Turbulence and turbulent transport}

Hydrodynamical turbulence, with its involved flow patterns 
of interwoven webs of eddies changing erratically in time
\cite{Richar}, is a fascinating phenomenon occurring in nature 
from microscopic scales to astronomical ones and exposed 
to our scrutiny in everyday experience. Despite the ubiquity 
of turbulence, an understanding of this far-from-equilibrium 
phenomenon from first principles is a long-standing open problem, 
one of the few remaining theoretical challenges of classical physics. 
Although it seems still a too far-fetched goal, there has been
a constant progress over years in the theoretical and practical knowledge 
of hydrodynamical turbulence and, even more so, of other non-equilibrium 
phenomena of a similar nature. This progress has been due to 
developments of experimental techniques and computer power, but 
also to new theoretical ideas inspired by studies of simple models 
of turbulence-related systems. The present lectures will discuss 
one circle of such ideas more relevant to the problem of transport 
properties in turbulent flows than to turbulence itself.
They concern properties of stochastic processes underlying the
turbulent mixing, see also \cite{SS,FGV}. 
In refs.\ \cite{MK,Warh,Saw,WiggOtt,TMGK} one may find discussion 
of other theoretical, experimental, numerical and practical aspects 
of turbulent transport. For the use of stochastic processes
in modeling of non-linear fluid dynamics itself, 
see e.g.\ \cite{ChevMan}.
\vskip 0.1cm

It is believed that realistic turbulent flows are described 
by hydrodynamical equations, like the Navier-Stokes ones, that govern 
the evolution of the velocity field and, eventually, of other relevant 
fields like fluid density, temperature, etc. 
As any structure whose dynamics is governed by evolution equations,
the turbulent flow may be viewed from a theoretical point as a 
dynamical system. Many attempts were made then to apply to turbulence 
ideas from the dynamical systems theory, see e.g.\ \cite{BJPV}. 
These ideas, developed for low-dimensional dynamical systems, although 
often successful in explaining the onset of turbulence, seem to miss 
important features when confronted with the fully developed turbulence. 
Some of them, however, prove very useful in the description of 
transport in turbulent flows, as we shall see in the sequel.

\subsection{Navier-Stokes equations}

These are equations dating back to the work of Claude-Louis Navier from 1823
and of George Gabriel Stokes from 1883, completing Leonhard Euler's 
equations of hydrodynamics from 251 years ago. They have the form of 
the Newton equation for the mass times acceleration of the the fluid element:
\begin{alignat*}{4}
& \rho (\partial_t \bm v + (\bm v \cdot \bm\nabla) \bm v) &
 \,-\,\rho & \nu \bm \nabla^2 \bm v
\ =\ &
  -\bm\nabla & p & \,+ & \bm f\,.
\\[-.6ex]
& \hspace{-1.4em}\nearrow &
& \hspace{-.3em}\uparrow &
& \hspace{-1.4em}\nearrow &
& \hspace{.3em}\nwarrow
\\[-.65ex]
& \hspace{-2em}{_\text{fluid}\atop^\text{density}} &
& \hspace{-3.2em}{_\text{kinematical}\atop^\text{viscosity}} &
& \hspace{-3em}^{\text{pressure}} &
& \hspace{1em}{_\text{force}\atop^\text{density}}
\end{alignat*}
and have to be supplemented with the continuity equation
$$\partial_t\rho+\bm\nabla\cdot(\rho\bm v)=0$$
and an equation of state $F(\rho,p)=0$. For the incompressible
fluid, the latter states that the density $\rho=\rho_0$ is constant
and the Navier-Stokes equations may be rewritten in the form 
\qq
\partial_t \bm v + (\bm v \cdot \bm\nabla) \bm v
 \,-\,\nu\bm\nabla^2\bm v
\ =\ {_1\over^{\rho_{_0}}}(-\bm\nabla p+\bm f)
\label{NS}
\qqq
and are supplemented with the incompressibility condition 
$\bm\nabla\cdot\bm v=0$. \,In the latter case, they
may be viewed as an evolution equation on the (infinite dimensional) 
space $\mathscr V_0$ of the divergence-free vector fields describing
an infinite-dimensional (non-autonomous) dynamical system
\qq
  \frac{\dd X}{\dd t}
=
  \mathscr{X}(t,X)
\qquad\text{for \ $X \in \mathscr V_0$.}
\qqq
In the theoretical modeling of the way in which the fluid motion is induced, 
one often assumes that the external force $\bm f$ is random.
The right hand side $\mathscr X(t,X)$ will then be also
random, describing an infinite-dimensional random dynamical system.
\vskip 0.1cm

Eq.\,(\ref{NS}) is nonlinear. The strength of the non-linear 
advection term $\,(\bm v\cdot\bm\nabla)\bm v\,$ relative to the linear one
$\nu\bm\nabla^2\bm v$ describing the viscous dissipation depends
on the length scale $\,l\,$ and is measured by the running Reynolds   
number defined as
$$
  \ReNo(l)
=
  \frac{\Delta_l v \cdot l}{\nu}
$$
where $\Delta_l v$ is the size of typical velocity difference
across the distance scale $\,l$, \,with $\,\tau_l=\frac{l}{\Delta_lv}\,$
giving the typical turnover time of eddies of size $\,l$.  
\,In particular, if $\,L\,$ is the integral scale corresponding 
to the size of the flow recipient, $\,\ReNo(L)\equiv\ReNo\,$ is the 
integral scale Reynolds number and $\,\tau_L\,$ is the integral time.
On the other end, $\,\ReNo(\eta) = 1\,$ for $\eta$ equal to the viscous 
(or Kolmogorov) scale on which the non-linear and the dissipative term 
in the equation have comparable strength. The Kolmogorov time 
$\,\tau_\eta\,$ gives the turnover time of the smallest eddies. 
Phenomenological observations of hydrodynamical flows point to the 
following classification according to the size of the Reynolds number 
$\ReNo$:
\begin{itemize} 
\item \quad\hbox to3.5cm{$\ReNo \lesssim 1$:\hfill}\text{laminar 
flows}, 
\item \quad\hbox to3.5cm{$\ReNo \sim 10
\ \text{to}\ 10^2$:\hfill}\text{onset of turbulence},
\item \quad\hbox to3.5cm{$\ReNo \gtrsim 10^3$:\hfill}\text{developed  
turbulence}. 
\end{itemize} 
In he laminar regime, some explicit solutions are known \cite{Batch}. 
At the onset of turbulence, the flow is driven by few unstable modes. 
The standard dynamical system theory studying temporal evolution 
of few degrees of freedom governed by ordinary differential 
equations or map iteration proved useful here e.g. in describing 
the scenarios of the appearance of chaotic motions, see e.g. \cite{Eck}. 
Finally, in the fully developed turbulence, there are many unstable 
degrees of freedom. Kolmogorov's scaling theory \cite{K41} for
this regime predicts that $$\Delta_lv\ \propto\ l^{1/3}
\qquad\text{for}\quad \eta\ll l\ll L$$ which is not very far from 
the observed behavior. The number of unstable modes may be estimated 
as given by $(L/\eta)^3\propto\ReNo^{9/4}$. New phenomena arise here
that have to be addressed, like cascades with (approximately) 
constant energy flux, intermittency, etc.

\subsection{Turbulent transport of particles and fields}

One may view a turbulent flow as a dynamical system in another way,
related to the transport of particles and fields by the fluid \cite{BJPV}.
   
\subsubsection{Transport of particles}

The particles may be idealized fluid elements, called \textbf{Lagrangian
particles}, or particles without inertia suspended in the fluid that 
also undergo molecular diffusion. We talk in the latter case of the 
\vskip 0.2cm

\begin{itemize}
\item\ \,\,\parbox[t]{12.9cm}{\textbf{tracer particles}
\vskip 0.1cm

whose motion is governed by the evolution equation
\qq
  \frac{\dd\bm R}{\dd t}
=
  \bm v(t,\bm R) + \sqrt{2\kappa}\, \bm\eta(t)\,,
\label{TP}
\qqq
where $\kappa$ stands for the (molecular) diffusivity and $\bm\eta(t)$ is 
the standard vector-valued white noise. The Lagrangian 
particles correspond to the case $\kappa=0$.}
\end{itemize}
\vskip 0.2cm

\noindent Small objects like water droplets in the air experience 
a friction force proportional to their relative velocity with respect 
to the fluid \cite{Max83}. These are 
\vskip 0.2cm

\begin{itemize}
\item\ \,\,\parbox[t]{12.9cm}{\textbf{particles with inertia}
\vskip 0.1cm

whose position $\bm R$ and velocity $\bm U$ satisfy in the simple
case the equations of motion
\qq
  \frac{\dd\bm R}{\dd t}
=
  \bm U
\ ,\quad
  \frac{\dd\bm U}{\dd t}
=
  \frac{_1}{^\tau}\big(-\bm U + \bm v(t,\bm R) + \sqrt{2\kappa}\, 
\bm\eta(t)\big)\,,
\label{IP}
\qqq
where $\tau$, the Stokes time, accounts for time delay with respect
to tracer particles whose evolution is recovered in the $\tau\to0$
limit.}
\end{itemize}
\vskip 0.2cm

\noindent Finally, one may consider small objects with spatial structure 
suspended in the fluid, like
\vskip 0.2cm

\begin{itemize}
\item\ \,\,\parbox[t]{12.9cm}{\textbf{polymer molecules}
\vskip 0.1cm

whose motion is sometimes modeled \cite{BHAC} by the differential equations 
\qq
  \frac{\dd\bm R}{\dd t}
=
  \bm v(t,\bm R) + \sqrt{2\kappa}\, \bm\eta(t)
,\ \quad
  \frac{\dd\bm B}{\dd t}
=
  (\bm B \cdot \bm\nabla) \bm v(t,\bm R)
 -\alpha \bm B + \sqrt{2\sigma}\, \bm\xi(t)\,,
\label{PC}
\qqq
where $\bm R$ describes the position of one end of the polymer chain,
$\bm B$ is the end-to-end separation vector, and $\bm\eta(t)$
and $\bm\xi(t)$ are independent white noises. The term $-\alpha\bm B$
describes the elastic counteraction to the stretching of the polymer.}
\end{itemize}
\vskip 0.2cm
   
\noindent All three cases provide examples of finite-dimensional
{\em dynamical systems} if $\bm v,\bm\eta,\bm\xi$ {\em are given}. 
The dynamical systems are random if $\bm v,\bm\eta,\bm\xi$ are random 
{\em with given statistics}. They may be studied with tools of the 
dynamical systems theory to which they have also provided new inputs.

\subsubsection{Transport of fields}

Turbulent flows may also transport physical quantities described by
fields changing continuously from point to point, like temperature,
pollutant or dye concentration, magnetic field, etc. The evolution
of such transported fields is described for small field intensity 
by linear advection-diffusion equations with variable coefficients 
related to the fluid velocity. Under the same assumption of small 
field intensity, one may ignore the influence of the transported 
field on the fluid dynamics, ending up with a {\em passive transport} 
approximation. The passive advection of a  
\vskip 0.2cm

\begin{itemize}
\item\ \,\,\parbox[t]{12.9cm}{\textbf{scalar field} $\,\theta(t,\bm r)$ 
\vskip 0.1cm

(e.g.\ temperature) is governed by the equation
\qq
  \partial_t \theta + (\bm v \cdot \bm\nabla)\theta
 -\kappa\bm\nabla^2 \theta
=
  g
\label{SF}
\qqq
where $\kappa$ stands for the diffusivity of the scalar and
$g$ is a scalar source.}
\end{itemize}
\vskip 0.2cm

\noindent Similarly, for a 
\vskip 0.2cm

\begin{itemize}
\item\ \,\,\parbox[t]{12.9cm}{\textbf{density field} $\,n(t,\bm r)$
\vskip 0.1cm

(e.g. of a pollutant), one has the partial differential equation
\qq
  \partial_t n + \bm\nabla \cdot (n \bm v )
 -\kappa\bm\nabla^2 n
=
  h
\label{SD}
\qqq
that differs from that for the scalar field only for compressible
velocities with $\bm\nabla \cdot \bm v \neq 0$.}
\end{itemize}
\vskip 0.2cm

\noindent For the passive transport of a
\vskip 0.2cm

\begin{itemize}
\item\ \,\,\parbox[t]{12.9cm}{\textbf{phase-space density} 
$\,n(t,\bm r,\bm u)$ 
\vskip 0.1cm

(e.g. of an aerosol suspension of inertial particles) one can write 
the equality
\qq
  \partial_t n + (\bm u \cdot \bm\nabla_{\bm r}) n+
  \bm\nabla_{\bm u} \cdot (n \frac{\bm u - \bm v}{\tau})
 -\kappa\bm\nabla^2 n
=
  h\,.
\label{PSD}
\qqq
}
\end{itemize}
\vskip -0.2cm

\noindent Finally, the passive transport of the 

\vskip 0.2cm
\begin{itemize}

\item\ \,\,\parbox[t]{12.9cm}{\textbf{magnetic field} 
$\,\bm B(t,\rm r)$
\vskip 0.1cm

satisfying $\bm\nabla\cdot\bm B=0$ is governed by the evolution equation
\qq
  \partial_t \bm B + (\bm v \cdot \bm\nabla) \bm B+(\bm\nabla\cdot\bm v)
  \bm B -(\bm B \cdot \nabla) \bm v - \kappa \bm\nabla^2 \bm B
=
  \bm G\,,
\label{MF}
\qqq
where $\kappa$ is the magnetic diffusivity and $\bm G$ is a source
term.}
\end{itemize}
\vskip 0.2cm

\noindent The meaning of  ``small intensity'' in the assumptions leading to 
the above equation depends on the physical situation (e.g.\ very small 
admixtures of polymers or chemical reactants may change the flow considerably).

\subsubsection{Relations between transport of particles 
and fields}

The dynamics of localized objects (particles, polymers) and passively 
transported fields listed above are closely related. 
\vskip 0.1cm

Let $\,\bm R(t;t_0,\bm r_0|\bm\eta)\,$ denote the time $t$ position 
of the tracer particle that at time $t_0$ passes through $\bm r_0$. 
\,It depends on the realization of the noise $\bm\eta$, see 
Eq.\,(\ref{TP}). Then the formula
\qq
\theta(t,\bm r)\,=\,\overline{\theta(t_0,\bm R(t_0;t,\bm r|\bm\eta))}\,=\,
\overline{\int\theta(t_0,\bm r_0)\,\,\delta(\bm r_0-\bm R(t_0;t,\bm r|\bm\eta))
\,\,\dd\bm r_0}\,,
\label{VCs}
\qqq
where the overline stands for the average over the white noise $\,\bm\eta\,$ 
realizations, expresses the solution of the advection-diffusion equation
(\ref{SF}) with vanishing source $\,g\,$ in terms of the
initial values $\,\theta(t_0,\bm r)\,$ of the field. The meaning of this
solution is that scalar field is conserved along the noisy trajectories. 
\vskip 0.1cm

Let us consider the matrix $\,W(t;t_0,\bm r_0|\bm\eta))\,$ that
propagates  the separations between two infinitesimally close 
tracer particle trajectories:
$$W^i_{\ j}(t;t_0,\bm r_0|\bm\eta)\ =\ \frac{\partial R^i(t;t_0,\bm
r_0|\bm\eta)}{\partial r_0^j}\,.$$ 
The determinant of matrix $\,W\,$ describes the flow-induced change 
of the volume 
element along the tracer particle trajectory and it is equal to $1$
for the incompressible flows. The formula
\qq
n(t,\bm r)\,=\,\overline{\det{W(t_0;t,\bm r|\bm\eta)}\,\,n(t_0,\bm R(t_0;
t,\bm r|\bm\eta))}\,=\,
\overline{\int n(t_0,\bm r_0)\,\delta(\bm r-\bm R(t;t_0,\bm r_0|\bm\eta))\,\,
\dd\bm r_0}
\label{VCd}
\qqq
gives the solution of the initial-value problem for the density
evolution equation (\ref{SD}) with vanishing source $\,h$. \,Note that
$$ W(t_0;t,\bm r|\bm\eta)\,=\,W(t;t_0,\bm R(t_0;t,\bm r)|\bm\eta)^{-1}$$
so that the density changes inversely proportionally to the change of volume
element along the Lagrangian trajectories. The density increases where 
the flow contracts the volume.
\vskip 0.1cm

Similarly, denoting by $\,\,(\bm R(t;t_0,\bm r_0,\bm u_0|\bm\eta),
\bm U(t;t_0,\bm r_0\,\bm u_0|\bm\eta))\,\,$ the phase-space trajectory 
of the inertial particle passing at time $\,t_0\,$ through  
$\,(\bm r_0,\bm u_0)\,$ and introducing the matrix 
$\,\bm W(t;t_0,\bm r_0,\bm u_0|\bm\eta)\,$ propagating the phase-space 
separations between two infinitesimally close inertial particle 
trajectories, one obtains the formula
\qq
n(t,\bm r,\bm u)\,=\,\overline{\det{\bm W(t_0;t,\bm r,\bm u|\bm\eta)}
\,\,n(t_0,\bm R(t_0;t,\bm r,\bm u|\bm\eta),\bm U(t_0;t,\bm r,\bm u|\bm\eta))} 
\qqq
for the solution of the evolution equation (\ref{PSD}) with $\,h=0$.
\vskip 0.1cm

Finally, the solution of the evolution equation (\ref{MF}) for the magnetic
field with $\,\bm G=0\,$ is given by the formula:
\qq
\bm B(t,\bm r)\,=\,\overline{(\det{W(t_0;t,\bm r|\bm\eta)})\,
\,W(t_0;t,\bm r|\bm\eta)^{-1}\,
\bm B(t_0,\bm R(t_0;t,\bm r|\bm\eta))}\,.
\qqq
The action of the matrix $\,\,W(t_0;t,\bm r|\bm\eta)^{-1}=W(t;t_0,\bm R(t_0;
t,\bm r|\bm\eta)|\bm\eta)\,\,$ stretches, contracts and/or rotates the
magnetic field vector along trajectories.
\,Note in passing that the solution of the second of Eqs.\,(\ref{PC}) for
the polymer end-to-end separation in the linear approximation
is given by
\qq
\bm B(t)\,=\,
  {\rm e}^{-\alpha(t-t_0)}\, W(t;t_0,\bm r_0|\bm\eta)\, \bm B(t_0)
\qqq
if the component $\,\sigma\,$ of the polymer diffusivity vanishes.
\vskip 0.1cm

The source terms providing inhomogeneous terms in the linear
advection-diffusion  equations for the fields may be taken 
into account by the standard ``variation of constants'' method. 
For example the solution (\ref{VCs}) of the scalar evolution
equation (\ref{SF}) picks up for non-zero source $\,g\,$ 
an additional term
\qq
\int\limits_{t_0}^t
\overline{g(s,\bm R(s;t,\bm r|\bm\eta))}\,ds 
\qqq
on the right hand side.

\vskip 0.5cm

\begin{conc}
Passive transport of particles and fields by hydrodynamical flows,
including turbulent ones, is described by simple equations closely 
related in both cases. The transport of particles will 
be studied below by the (random) dynamical system methods. 
We shall subsequently examine which attributes of the particle dynamics 
bear on which properties of the field advection, with the goal to
capture and explain some essential features of the turbulent transport.  
\end{conc}

\newpage

\nsection{Multiplicative ergodic theory}

We shall be interested in a statistical description of particle 
dynamics in turbulent flows. For concreteness, we shall concentrate
on the dynamics of Lagrangian particles carried by the a flow in a bounded 
region $V$. Their trajectories are solutions of the ODE 
\eq{\label{Lagr-traj}
  \frac{\dd \bm R}{\dd t}
=
 \bm v(t,\bm R)\,.
}
with random, not necessarily incompressible, velocity fields 
$\,\bm v(t,\bm r)$, \,see Eq.\,(\ref{TP}).
\,We shall assume that the statistics of the
velocities is stationary. The deterministic time-independent velocity 
field will be also covered as a special case. 
Since Eq.\,(\ref{Lagr-traj}) has a form of a general dynamical
system, most of the considerations that follow will apply
without much change also to other particle-transport equations.
For reasons which were already mentioned in the introduction 
and will become clear in later, the considerations of the present
and the next lecture, borrowing on the theory of differentiable 
dynamical systems, are relevant to the advection by turbulent 
flows at intermediate Reynolds numbers.

\subsection{Natural invariant measures}

In deterministic time-independent flows, one calls a measure $\,n(\dd\bm r)\,$
on the flow region $\,V\,$ invariant if
\qq
\int\limits_V f(\bm R(t;\bm r)\,n(\dd\bm r)\,=\,\int\limits_V 
f(\bm r)\,n(\dd\bm r)
\qqq
for bounded (measurable) functions $\,f\,$ on $\,V\,$ and all times
$\,t$, with the shortened notation $\,\bm R(t;\bm r)\equiv\bm R(t;0,\bm r)$. 
\,This may be generalized to the case of stationary random flows 
by considering a collection of measures $\,n(\dd\bm r|\bm v)\,$ on the flow
region, parametrized by the velocity realizations, such that
\qq
\Big\langle \int\limits_V f(\bm R(t;\bm r|\bm v)|\bm v_t)\,
\,n(\dd\bm r|\bm v)\Big\rangle\,=\,\Big\langle \int\limits_V 
f(\bm r|\bm v)\,\,n(\dd\bm r|\bm v)\Big\rangle\ \equiv\ 
\int\limits_{V\times\CV} f(\bm r|\bm v)\,\,N(\dd\bm r|\dd\bm v)
\label{invm}
\qqq
where $\,\big\langle\,-\,\big\rangle\,$ denotes the average over the 
velocity ensemble and $\,\bm v_t(s,\bm r)=\bm v(s+t,\bm r)$.
\,Such a collection defines a measure $\,N(d\bm r|d\bm v)\,$ on the
product $\,V\times\CV\,$ of the flow region $\,V\,$ and the space
$\,\CV\,$ of velocity realizations. This measure is invariant under 
the 1-parameter group of transformations
$$V\times\CV\,\ni\,(\bm r|\bm v)\ \longmapsto\ (\bm R(t;\bm r|\bm v),
\m\bm v)\,.$$
It is called an invariant measure of the random dynamical system 
(\ref{Lagr-traj}).
\vskip 0.1cm

Let us seed the particles into the fluid at some past time $\,t_0\,$ 
with smooth normalized density $n_{t_0}(t_0,\bm r_0)$, \,e.g. the constant 
one. This density will evolve in time according to the advection equation
$\,\partial_t n+\bm\nabla\cdot(n\bm v)=0$, \,see Eq.\,(\ref{SD}).
\,In an incompressible flow, the constant density $\,n_{t_0}(t,\bm r)
=|V|^{-1}\,$ will stay constant and normalized also at all later times 
$\,t$. \,If the flow is compressible, however, then the particles will 
develop preferential concentrations and the densities $n_{t_0}(t,\bm r|\bm v)$ 
will become rougher and rougher with time in typical velocity realizations
$\,\bm v$, as illustrated by the snapshots taken form \cite{BDES}:

\begin{figure}[h!]
\begin{center}
\vspace{0.1cm}
\mbox{\hspace{0.0cm}\psfig{file=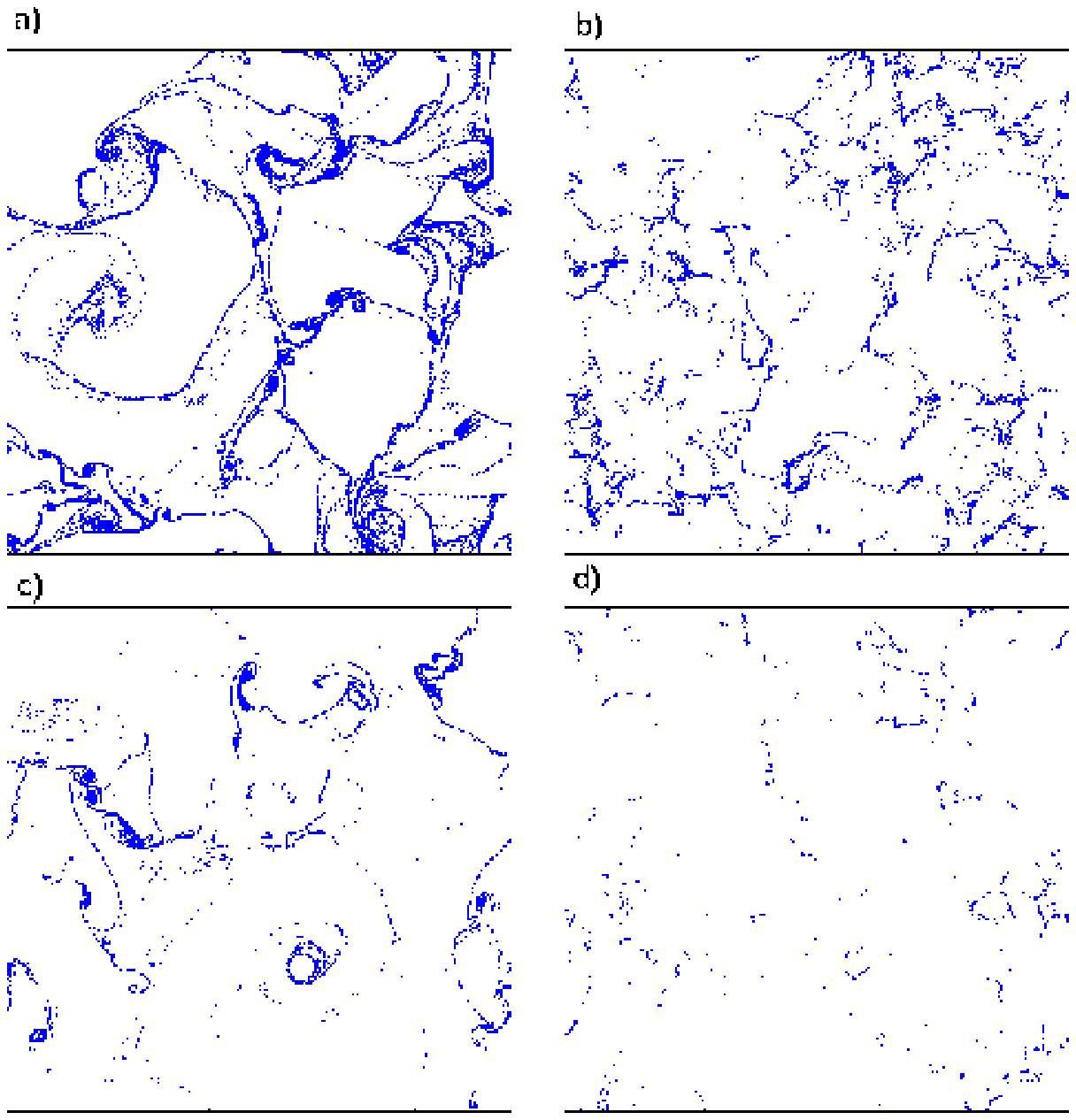,height=4.5cm,width=4.5cm}}
\end{center}
\end{figure}

\noindent Similar phenomenon occurs for the phase-space density 
of inertial particles, see e.g.\ \cite{Bec,BBCLMT}. \,The probability measures 
$\,n(\dd \bm r| \bm v)\,$ on $\,V\,$ are called 
natural measures if
$$
  \int\limits_V f(\bm r)\,n(t,\dd\bm r|\bm v)
=
  \lim_{t_0\to-\infty} \frac{1}{|t_0|}
    \int_{t_0}^0\int\limits_V f(\bm r)\,n_{t_0}(0,\dd\bm r|\bm v)
$$
for all continuous functions $\,f\,$ and almost all velocity realizations
$\,\bm v$. \,In particular, $\,\int f(\bm r)\,n(\dd\bm r|\bm v)
=\lim\limits_{t_0\to-\infty}
\int f(\bm r)\,n_{t_0}(0,\bm r|\bm v)\,d\bm r\,$ if the latter limits exist.
In that case, the natural measures are just the time zero distributions
of particles seeded with a smooth density at time $-\infty$.
For compressible flows, the natural measures $n(\dd\bm r|\bm v)$ are 
concentrated on a random time- and velocity-field-dependent attractor. 
These measures have the property (\ref{invm}) and give rise to the
natural invariant measure $\,N(\dd\bm r|\dd\bm v)$.
The latter permits to do statistics of the values of functions
$\,f(\bm r|\bm v)\,$ of points in the flow region and velocity realizations.
Such statistics is often called ``Lagrangian'' since it may be calculated
from time-averages along the Lagrangian trajectories.

\subsection{Tangent flow}

Recall that the matrix with the elements $$W^i_{\ j}(t;t_0,\bm r|\bm v) 
\,=\,\frac{\partial R^i(t;t_0,\bm r|\bm v)}{\partial r^j}$$ propagates 
infinitesimal separations $\delta\bm R$ between Lagrangian trajectories
in a given velocity realization:
$$\delta\bm R(t;t_0,\bm r|\bm v)\,=\,W(t;t_0,\bm r|\bm v)\,
\delta\bm r\,.$$  
We shall abbreviate $\,W(t;0,\bm r|\bm v) \equiv W(t;\bm r|\bm v)\,$ 
dropping also the other dependences in $\,W\,$ from the notation whenever 
they are not necessary. Diagonalizing the matrix $\,W^TW\,$ and writing its
positive eigenvalues in the exponential form as
$\,\ee^{\m2\rho_1}\geq\ee^{\m2\rho_2}\geq\,\cdots\,\geq\ee^{\m2\rho_d}$,
\,we obtain the so called ``stretching exponents'' 
$\,\rho_i(t;\bm r|\bm v)\,$ with the ordering
$$\rho_1\,\geq\rho_2\,\geq\,\cdots\,\geq\rho_d\,.$$
Due to the cumulative effect of short-time stretchings, contractions 
and rotations of the separation vectors of infinitesimally closed 
Lagrangian trajectories, the stretching exponents typically grow in absolute
value in time. We shall be interested in their long-time asymptotics. 
Let us call the ratios $\,\sigma_i=\frac{\rho_i}{|t|}\,$ the ``stretching 
rates''. \,They are sometimes also called ``finite-time Lyapunov exponents''.
For fixed time $\,t$, \,the functions $\,\sigma_i(t;\bm r|\bm v)\,$ may 
be treated as random variables on the product space $\,V\times\CV\,$ 
equipped with the natural invariant measure $\,N(\dd\bm r|\dd\bm v)$. 
\,Their joint probability density function (PDF) is then given by 
the formula:  
$$
  P_t(\vec\sigma)
=
  \int\limits_{V \times \mathscr V}\hspace{-0.1cm} 
    \delta(\vec\sigma - \vec\sigma(t;\bm r, \bm v))\,
    N(\dd\bm r| \dd\bm v)\,.
$$
The PDF's $\,P_t(\vec{\sigma})\,$ and $\,P_{-t}(\vec{\sigma})\,$ are 
very simply related:
\qq
P_t(\vec\sigma)\,=\,P_{-t}(-\antivec\sigma)\,,
\qqq
where $\,\antivec\sigma=(\sigma_d,\dotsc,\sigma_1)\,$ if 
$\,\vec\sigma =(\sigma_1,\dotsc,\sigma_d)$.
\,This follows from the relation 
$\,\bm R(-t;\bm R(t;\bm r|\bm v)|\bm v_t)=\bm r\,$ implying
by the chain rule that $\,\,W(t;\bm r|\bm v)=
W(-t;\bm R(t,\bm r|\bm v)|\bm v_t)^{-1}\,\,$ and that
\qq
\vec\sigma(t;\bm r|\bm v)\,=\,
-\antivec\sigma(-t;\bm R(t,\bm r| \bm v)|\bm v_t)
\label{rrss}
\qqq 
and from the invariance (\ref{invm}) of the measure 
$\,N(\dd\bm r|\dd\bm v)$.
\,A very general result, the backbone of the multiplicative ergodic theory,
assures that the stretching rates approach constant limiting
values for long times. Denote $\ln_+ x = \max(0,\ln x)$.  
\vskip 0.3cm

\begin{itemize}
\item{\bf Multiplicative Ergodic Theorem}\ \,(Oseledec 1968 \cite{Osel}, 
Ruelle 1979 \cite{Ruelle1}). 
\vskip 0.1cm

\noindent If the measure invariant measure $\,N(\dd\bm r|\dd\bm v)\,$ 
is ergodic and $\int \ln_+ \| W(t;\bm r|\bm v)\| N(\dd\bm r, \dd\bm v) <
\infty$ for some $t>0$ then
\qq
\label{MET}
  \lim_{t\to\infty}\ P_t(\vec\sigma)\ =\ \delta(\vec\sigma-\vec\lambda)
\qqq
for a vector $\,\vec\lambda=(\lambda_1,\dots,\lambda_d)\,$ of 
``Lyapunov exponents'' such that $\,\lambda_1\geq\cdots\geq\lambda_d$.
\end{itemize}
\vskip 0.2cm

\noindent This may be viewed as a particular case of a multiplicative 
version of the law of large numbers \cite{LArnold}. \,For typical $\,\bm r$, 
$\,\bm v\,$ and $\,\delta\bm r$, 
$$
  \lambda_1
\,=\,
  \lim_{t\to\infty}\ \frac{1}{t}\, \ln \| W(t;\bm r|\bm v)\,\delta\bm r\|\,.
$$
The strict positivity of $\lambda_1$ signals the sensitive dependence 
on the initial conditions and is taken as a definition of ``chaos''. 
\,Similarly, for $1 \leq n \leq d$ and typical 
$\,\delta\bm r_1,\dots,\delta\bm r_n$,
\qq
  \lambda_1 + \dotsb + \lambda_n\,=\,
  \lim_{t\to\infty}\,\, \frac{1}{2t}\,
    \ln \det \Bigl(
      \delta\bm r_i \scprod
       W^TW(t;\bm r|\bm v)\, \delta\bm r_j)_{1\leq i,j \leq n}
    \Bigr).
\qqq

\noindent It is important to know how the limit (\ref{MET})
is approached. \,If $\,\lambda_1 > \dotsb > \lambda_d\,$ then one may expect
that the fluctuations of the stretching rates about their limiting values
behave according to
\vskip 0.2cm

\begin{itemize}
\item{\bf Multiplicative central limit} {\em\,(not covered by a general 
theorem)} \,asserting that 
$$
  \lim\limits_{t\to\infty}\ t^{-1/2} P_t(\lambda + t^{-1/2}\vec\tau)
\ =\ \frac{e^{-\frac12 \vec\tau \scprod C^{-1}\vec\tau}}
       {\det(2\pi C)^{1/2}}\,.\quad
$$
\end{itemize}
\vskip 0.1cm

\noindent Finally, as argued in \cite{BalkFoux}, see also
\cite{CDG}, again under the assumption that $\,\lambda_1 > \dotsb > \lambda_d$,
\,one may expect occurrence of the regime of
\vskip 0.3cm

\begin{itemize}
\item{\bf Multiplicative large deviations} {\em\,(again not covered
by a general theorem)}, \,where
$$
  -\,\lim\limits_{t\to\infty}\ \frac{_1}{^t}\,\ln{P_t(\vec\sigma)}
\ =\ 
  H(\vec\sigma)\quad
$$
for a ``rate function'' $\,H\geq0\,$ such that $\,H(\vec\lambda)=0$.
\end{itemize}
\vskip 0.2cm

\noindent For incompressible velocities where $\,\sum\sigma_i
\equiv 0$, \,the rate function $\,H(\vec\sigma)\,$ is infinite 
unless $\,\vec\sigma\,$ satisfies the latter condition.
Loosely speaking, the existence of the regime of large deviations 
means that $$P_t(\vec\sigma)\ \ \approx\ \ \ee^{-tH(\vec\sigma)}$$
for large $\,t$. \,As in the additive case, the existence of such regime
with the rate function $\,H(\vec\sigma)\,$ twice differentiable around
the minimum implies the central limit behavior with 
$$
  (C^{-1})^{ij}
=
  \frac{\partial^2 H}{\partial\sigma_i \partial\sigma_j} (\vec\lambda)
$$
There are partial results about existence of the multiplicative large
deviation regime for deterministic dynamical systems \cite{Grass} and 
for random dynamical systems with decorrelated velocities
\cite{BaxStr,CDG}. The latter case will be discussed in the next lecture.
The large deviation regime for the stretching rates in realistic flows 
becomes accessible numerically and even in experiments \cite{BDD,BCG}. 
\vskip 0.1cm

Let us denote by $\,\bm v'\,$ the time-reversed velocity field:
$\,\bm v'(t,\bm r)=-\bm v(t,\bm r)$.  \,The large deviations rate 
function $\,H(\vec\sigma)\,$ and $\,H'(\vec\sigma)$, \, the latter one
pertaining to the flow in the time-reversed velocities, should satisfy
\vskip 0.2cm

\begin{itemize}
\item{\bf Multiplicative fluctuation relation} \,(of Gallavotti-Cohen 
type)
\qq
\label{MFR}
  H'(-\antivec\sigma)
=
  H(\vec\sigma) - \sum_{i=1}^d \sigma_i
\qqq
\end{itemize}
\vskip 0.1cm

\noindent In particular, for time-reversible velocities when the fields
$\,\bm v\,$ and $\,\bm v'\,$ have the same distribution, the two rate
function coincide: $\,H'=H$. \,The original fluctuation relations 
studied the large deviation regime of $\,s = -\sum_{i=1}^d \rho_i\,$ 
equal to the (phase-)space contraction exponent. The latter has
been interpreted \cite{Ruelle2,Ruelle3} as the entropy
production of the dynamical system, with $\,\sigma = s/t\,$ standing
for the entropy production rate. For deterministic uniformly hyperbolic 
(discrete-time) systems that are time-reversible, Gallavotti and Cohen 
\cite{GalCoh} established that the rate function $\,h(\sigma)\,$ of 
large deviations of $\,s\,$ satisfies the relation
$$
\label{FR}
  h(-\sigma)
=
  h(\sigma) + \sigma\,.$$
This was generalized recently in \cite{BonGalGent} to random uniformly
hyperbolic systems.  Since
$$
  h(\sigma)
=
  \min_{-\sum \sigma_i = \sigma} H(\vec\sigma)
\,,
$$
the relation (\ref{MFR}) implies Eq.\,(\ref{FR}).
\vskip 0.2cm

Let us prove here following \cite{BalkFalkFoux}, see also \cite{FGV}, 
another, simpler, relation for a modified joint PDF of the time $\,t\,$ 
stretching rates defined by
\qq
  \tilde P_t(\vec\sigma)
=
  \left\langle
    \int \delta(\vec\sigma - \vec\sigma(t;\bm r|\bm v))\,\frac{\dd\bm r}{|V|}
  \right\rangle
\label{tildpt}
\qqq
Note that $\tilde P_t$ employs $\frac{\dd\bm r}{|V|}$ to average over $\bm r$
rather than the natural measures $n(\dd\bm r|\bm v)$ used in $P_t$. \,
Numerically, it is in general simpler to attain the PDF $\,\tilde P_t\,$
than $\,P_t\,$ as the latter requires performing longer Lagrangian averages.
\vskip 0.2cm

\begin{itemize} 
\item{\bf Transient multiplicative fluctuation relation}\,\,
(of the Evans-Searls type)
\qq
\label{TMFR}
  \tilde P'_{t}(-\antivec\sigma)
\,=\,
  \tilde P_t(\vec\sigma)\,\,\ee^{\m|t|\sum_i \sigma_i}\,,
\qqq
where $\,\tilde P'(\vec\sigma)\,$ pertains to the time-reversed flow.
\end{itemize}
\vskip 0.1cm

\noindent The proof of the relation (\ref{TMFR}) employs a simple 
change-of-variables argument similar to the ones used to obtain transient 
fluctuation relations \cite{EvSear1,EvSear2,Jarz1}.
First, using the definition \eqref{tildpt}
and the stationarity of the velocity ensemble, we obtain
\qq
  \tilde P_{-t}(-\antivec\sigma)
\,=\,
  \left\langle
    \int \delta(-\antivec\sigma - \vec\sigma(-t;\bm R|\bm v))\,
    \frac{\dd\bm R}{|V|}
  \right\rangle
\,=\,
  \left\langle
    \int \delta(\vec\sigma + \antivec\sigma(-t;\bm R|\bm v_t))\,
    \frac{\dd\bm R}{|V|}
  \right\rangle.
\qqq
Upon the substitution $\,\bm R=\bm R(t; \bm r|\bm v)\,$ with the Jacobian
\qq
  \frac{\partial(\bm R)}{\partial(\bm r)}
\,=\,
  \det W(t; \bm r|\bm v)
\,=\,
  e^{\m|t|\sum_{i=1}^d \sigma_i(t;\bm r|\bm v)}
\qqq
this gives
\qq
  \tilde P_{-t}(-\antivec\sigma)
&=&
  \left\langle
    \int \delta(\vec\sigma + \antivec\sigma(-t;\bm R(t; \bm r|
    \bm v)|\bm v_t))\,\,
    \ee^{\m|t|\sum_{i=1}^d \sigma_i(t;\bm r|\bm v)}\,
    \frac{\dd\bm r}{|V|}
  \right\rangle\cr
&=&
  \left\langle
    \int \delta(\vec\sigma - \vec\sigma(t;\bm r|\bm v))
    \,\,\ee^{\m|t|\sum_{i=1}^d \sigma_i(t;\bm r|\bm v)}\,
    \frac{\dd\bm r}{|V|}
  \right\rangle
\ =\ 
  \tilde P_t(\vec\sigma)\,\,\ee^{\m|t|\sum_{i=1}^d \sigma_i}
\nonumber
\qqq
where we have use the relation \eqref{rrss} to obtain the second equality.
Now, the straightforward relation $\,R(-t;\bm r|\bm v)=R(t;\bm r|\bm v')\,$ 
implies that
\qq
\tilde P_{-t}(\vec\sigma)\,=\,\tilde P'_t(\vec\sigma)\,.
\qqq
and the identity (\ref{TMFR}) follows.
\vskip 0.1cm

For long positive times $\,t\gg0$,  \,the Lagrangian
particles approach the (time-varying) attractor and 
their statistics should not depend much on whether their initial
points were distributed with the uniform measure
or with the attractor measure. One may then expect
that
\qq
  \tilde P_t(\vec\sigma)
\ \approx\ 
  P_t(\vec\sigma)
\label{appp}
\qqq
at long positive times, and similarly for $\,\tilde P'_t\,$ and $\,P'_t$.
\,If this holds and the PDF's $\,P_t\,$ and $\,P'_t\,$ exhibit 
large deviations regime then \eqref{TMFR} implies \eqref{MFR}. 
Both assumptions are not granted, however, and require a careful
analysis in specific situations.

\subsection{Applications of multiplicative large deviations}
\label{sec:uses}

Although concerning relatively rare events, large deviations play 
an important role in turbulent transport phenomena. Indeed, rare 
fluctuations of concentrations of toxic emissions or chemical agents
may determine the nocivity and/or the reaction rates. Suppose that 
we put at time zero a blob of density $\,n_0(\bm r_0)\,$ into
a compressible fluid. Then, for vanishing diffusivity\footnote{At
long distances and not too long times, the effect of molecular
diffusivity my be disregarded in practical situations.}, the blob 
density at later times is 
\qq
  n(t,\bm r)\,=\,\det{W(0;t,\bm r)}\,\,n_0(\bm R(0;t\bm r)\,,
\qqq
see Eq.\,(\ref{VCd}). Along the trajectory $\,\bm R(t;\bm r_0)$,
$$
  n(t)\ 
\equiv\ 
  n(t,\bm R(t;\bm r_0))
\,=\,
  \det W(t;\bm r_0)^{-1} n_0(\bm r_0)
\,=\,
  \ee^{-\sum_i \rho_i(t;\bm r_0)} n_0(\bm r_0)
$$
and
\qq
  \big\langle n(t)^q\big\rangle
\,=\,
  n_0(\bm r_0)^q
  \int\ee^{-q t \sum\sigma_i}\,\tilde P_t(\vec\sigma) \,\dd\vec\sigma
\ \approx\ 
  n_0(\bm r_0)^q
 \int\ee^{-q t\sum \sigma_i - t H(\vec\sigma)} \,\dd\vec\sigma
\ \ \propto\ \ 
  \ee^{\m\gamma_q t}
\qqq
for large $\,t$, \,where 
\qq
-\gamma_q\,=\, \min\limits_{\sigma_1\geq\dotsb\geq\sigma_d}\,
q\sum_i \sigma_i + H(\vec\sigma)\,.
\label{galp}
\qqq
As we see, the asymptotic behavior of the moments of the density
of the blob along a Lagrangian trajectory is determined by
the rate function $\,H(\vec\sigma)$. 
\,Of course, $\,\gamma_0=0$. \,Eq.\,(\ref{MFR}) implies, in turn, 
that $\,\gamma_{-1}=0\,$ since the minimum of 
$\,H'(-\antivec{\sigma})\,$ is equal to zero. In the incompressible
flow, all $\,\gamma_q\,$ vanish.  
\vskip 0.1cm

There are other quantities that may be expressed via the rate function
$H(\vec\sigma)$ of the large deviations of the stretching exponents including:
\begin{itemize}
\item rates of decay of moments of the tracer in the presence of diffusivity
  \cite{BalkFoux}
\item multifractal dimensions of the natural attractor measure
  \cite{Grass,BGH}
\item evolution of the moments of the polymer stretching $\bm B$ and
  the onset of drag reduction in polymer solutions
  \cite{BalkFouxLeb,Chertkov}
\end{itemize}
\vskip 0.3cm

\begin{conc}
In smooth dynamical systems cumulation of local stretchings, contractions
and rotations leads to the linear growth of the stretching exponents
with the asymptotic rates equal to the Lyapunov exponents. If the
Lyapunov exponents are all different then one expects that the statistics
of the stretching exponents exhibits at long but finite times 
a large deviations regime characterized by the rate function
$\,H(\vec\sigma)$. Such a function contains more information about
the short-distance long-time properties of the flow than the
Lyapunov exponents and it enters the determination of several asymptotic 
transport properties of turbulent velocities at intermediate Reynolds
numbers. The rate functions $\,H\,$ for the direct and time-reversed flows 
are related by the Gallavotti-Cohen type symmetry (\ref{MFR}).  
\end{conc}

\newpage

\nsection{Particles in smooth Kraichnan velocities}

In 1968 Robert~H.~Kraichnan initiated the study of turbulent
transport in a Gaussian ensemble of velocities decorrelated in time.  
The statistics of such velocities is determined by the mean
$\,\langle v^i(t,\bm r) \rangle$, \,that we shall take equal to zero, 
\,and by the covariance of the form
$$
  \big\langle v^i(t,\bm r)\,v^j(t',\bm r') \big\rangle
\,=\,
  \delta(t-t')\,D^{ij}(\bm r,\bm r')\,.
$$
To imitate turbulent flows far from boundaries, one may assume:
\begin{itemize}
\item\quad\hbox to 3cm{homogeneity\hfill}$\,D^{ij}(\bm r,\bm r')\,
=\,D^{ij}(\bm r - \bm r')\,$,
\item\quad\hbox to 3cm{isotropy\hfill}$\,D^{ij}(\bm r)\,=\,
  \delta^{ij}\,D_1(|\bm r|)
 \,+\,{r^i r^j}\,D_2(|\bm r|)\,$,
\item\quad\hbox to 3cm{scaling\hfill}$\,D^{ij}(\bm r)-D^{ij}(\bm0)\,=\,
  \begin{cases}
    O(\bm r^2) & \text{for $|\bm r| \ll \eta$,} \\
    O(|\bm r|^\xi) & \text{for $\eta \ll |\bm r| \ll L$.}
  \end{cases}$
\end{itemize}

\noindent The range where $\,D^{ij}(\bm r)-D^{ij}(\bm0)\,$ is approximately 
quadratic is called the ``Batchelor regime'' and is used to mimic flows
at intermediate Reynolds numbers, dominated by viscous effects, with 
the spatial velocity increments approximately linear in the point separation.
The range where $\,D^{ij}(\bm r)-D^{ij}(\bm0)\,$ scales with power 
$\xi$ between 0 and 2 mimics the developed ``inertial range'' turbulence
at high Reynolds numbers, where the spatial velocity increments scale 
like a fractional power of the distance between the points (the Kolmogorov 
scaling is rendered by $\,\xi=4/3$). \,Although not very realistic, 
the Kraichnan model of turbulent velocities has provided a very useful 
playground for theoretical and numerical studies of transport properties 
in random flows. It allowed to test old ideas about turbulent transport and,
even more importantly, to come up with important new ideas \cite{FGV}.

\subsection{Lagrangian particles and tangent flow in Kraichnan
model}

In the Kraichnan ensemble, velocities are white noise (so
distributional) in time and the Lagrangian trajectories are defined by
the stochastic differential equation (SDE) 
$$
  \dd\bm R
=
  \bm v(t,\bm R) \,\dd t
\label{LSDE}
$$
that we shall interpret with the It\^o convention \cite{Risk,Oksen}, 
i.e.\ as equivalent to the integral equation
$$
  \bm R(t)
\,=\,
  \bm R(t_0)
 \,+\,
    \lim_{t_0<t_1<\dotsb<t_n<t}
    \sum_i \int_{t_i}^{t_{i+1}} \bm v(s, \bm R(t_i)) \,\dd s\,,
  $$
where the limit of the Riemann sums defines the It\^o stochastic integral
$\,\int_{t_0}^t\! \bm v(s,\bm R(s)) \,\dd s$. \,Note that the average
value of the It\^o integral vanishes since $\,\langle \bm v(s,\bm R(t_i)
\rangle=0\,$ for $\,s\geq t_i$. 
\,The Stratonovich convention would use the Riemann sums with
$\,\bm v(s,\bm R(t_i))\,$ replaced by $\bm v(s,
\bm R(\frac{_1}{^2}(t_i+t_{i+1})))\,$ but it gives the same result if 
$\,\partial_i D^{ij}(\bm r,\bm r)\equiv0$, \,which holds 
e.g.\ for homogeneous and isotropic velocities. 
\vskip 0.1cm
 
For a regular function $\,f$, \,the It\^o stochastic calculus gives the SDE
for the composed process with an additional (It\^o) term as compared
to the standard chain rule: 
$$
  \dd f(\bm R)
\,=\,
  \bm v(t,\bm R) \scprod \bm\nabla f(\bm R) \,\dd t
 +\half D^{ij}(\bm 0) \nabla_i \nabla_j f(\bm R) \,\dd t
$$
with the assumption of homogeneity. \,For the averages, one obtains 
the differential equation:
$$
  \frac{\dd}{\dd t} \big\langle f(\bm R) \big\rangle
\,=\,
  \big\langle \half D^{ij}(0) \nabla_i \nabla_j f(\bm R)\big\rangle
$$
which may be rewritten in the form
$$
  \frac{\dd}{\dd t}
  \int f(\bm r)\,\big\langle\delta(\bm r-\bm R(t;t_0,\bm r_0))\big\rangle
  \,\dd\bm r
\,=\,
  \int \half D^{ij}(\bm 0)\,f(\bm r)
    \,\big\langle \delta(\bm r - \bm R(t;t_0,\bm r_0))\big\rangle
  \,\dd\bm r
$$
or, stripping the last identity from arbitrary functions $\,f$, \,as
$$
  \frac{\dd}{\dd t}
  \big\langle\delta(\bm r-\bm R(t;t_0,\bm r_0))\big\rangle
\,=\,
  \half D^{ij}(\bm 0)\,\nabla_{r^i} \nabla_{r^j}
    \big\langle \delta(\bm r - \bm R(t;t_0,\bm r_0))\big\rangle\,.
$$
With the additional assumption of isotropy, 
$\,\frac{1}{2}D^{ij}(\bm 0) = D_0\m\delta^{ij}$, \,and we 
infer that the probability to find the Lagrangian trajectory at time 
$\,t\,$ at point $\,\bm r\,$ is that of the diffusing particle:
$$
  P(t_0,r_0;t,r)
\,\equiv\,
  \big\langle\delta(\bm r-\bm R(t;t_0,\bm r_0))\big\rangle
\,=\,
  \ee^{|t-t_0| D_0 \bm\nabla^2}\hspace{-0.07cm}(\bm r_0,\bm r)
\,=\,
  \frac{1}{(4\pi D_0|t-t_0|)^{d/2}}
  \ee^{-\frac{(\bm r-\bm r_0)^2}{4D_0|t-t_0|}}\,.
$$
The constant $\,D_0\,$ is called the ``eddy diffusivity''.
One infers that in mean, the Kraichnan turbulence causes diffusion.  
In particular, for the mean density and the mean scalar,
\qq
  \big\langle n(t,\bm r)\big\rangle
&=&
  \big\langle
    \int n(t_0,\bm r_0)\,\,\delta(\bm r-\bm R(t;t_0,\bm r_0))
    \,\,\dd\bm r_0
  \big\rangle
\ =\ 
  \int n(t_0,\bm r_0)\,\,\ee^{\m(t-t_0)D_0\bm\nabla^2}\hspace{-0.07cm} (\bm r_0,\bm r)\,\,\dd\bm r_0\,,\cr
\big\langle \theta(t,\bm r)\big\rangle
&=&
  \big\langle \int \theta(t_0,\bm r_0)\,\,\delta(\bm r_0-\bm R(t_0;t,\bm r))
  \,\,\dd\bm r_0\big\rangle
\ =\ 
  \int \theta(t_0,\bm r_0)\,\,\ee^{\m(t-t_0)D_0\bm\nabla^2}\hspace{-0.07cm} 
(\bm r,\bm r_0)\,\,\dd\bm r_0
\nonumber
\qqq
for $\,t \geq t_0$, \,see Eqs.\,(\ref{VCd}) and (\ref{VCs}).
An initial blob of scalar or density diffuses in mean as watched in
the laboratory frame. Of coarse, the evolution of the average density 
does not contain information about the dynamical behavior of the density 
fluctuations. A way to catch a glimpse of the latter is to look 
at the evolution of the blob from the point of view that moves with 
the fluid, as was discussed in Sect.\,\ref{sec:uses}. 
\vskip 0.1cm

To this end, let us study the dynamics of the small 
(infinitesimal) separation between 
the Lagrangian trajectories that satisfies the SDE
\qq
  \dd\delta\bm R
\,=\,
  (\delta\bm R \scprod \bm\nabla \bm v)(t,\bm R(t)) \,\dd t
\,\equiv\,
  \Sigma(t)\, \delta\bm R \,\dd t\,,
\label{Sigma}
\qqq
where $\,{\Sigma^i}_j(t) = (\nabla_j v^i)(t,\bm R(t))$.  \,Again, we shall
consider this equation with the It\^o convention but the Stratonovich
convention would give again the same result under the assumption that
$\,\partial_i D^{ij}(\bm r,\bm r)\equiv0$. \,For a function of $\,\delta\bm
R\,$, \,the It\^o calculus gives:
$$
  \dd f(\delta\bm R)
\,=\,
  (\delta\bm R \scprod \bm\nabla
    \bm v)(t,\bm R(t)) \scprod \nabla f(\delta\bm R) \,\dd t
 \,-\,\half \delta R^k \delta R^l\,(\nabla_k \nabla_l
    D^{ij})(\bm0)\, (\nabla_i \nabla_j f)(\delta\bm R) \,\dd t
$$
in the homogeneous case and for its average,
\qq
  \frac{\dd}{\dd t} \big\langle f(\delta\bm R) \big\rangle
\,=\,
  \big\langle
    \half \delta R^k \delta R^l\,C^{ij}_{kl}\,
    (\nabla_i \nabla_j f)(\delta\bm R)
  \big\rangle\,,
\label{fita}
\qqq
where $\,C^{ij}_{kl} = -\nabla_k \nabla_l D^{ij}(0)$.
\,The last result is the same as if $\,\delta\bm R(t)\,$ satisfied the
linear It\^o SDE
\qq
\label{eq:star2}
  \dd\delta\bm R
\,=\,
  S(t)\,\delta\bm R \,\dd t
\qqq
with the matrix-valued white noise $\,S(t)\,$ such that $\,\langle S(t)
\rangle = 0\,$ and $\,\langle {S^i}_k(t)\,{S^j}_l(t') \rangle = C^{ij}_{kl}
\,\delta(t-t')$.  \,Here taking the It\^o prescription is essential
since although the It\^o and Stratonovich prescriptions give the same
result for Eq.\,(\ref{Sigma}), they do
not agree for Eq.\,(\ref{eq:star2}) with $\,{S^i}_j(t)\,$ being 
the white noise! \,Similarly, the statistics of the tangent process 
$\,W(t;\bm r_0)\,$ for a fixed initial point $\,\bm r_0\,$ may 
be obtained by solving the linear It\^o SDE
\qq
\label{eq:star2b}
  \dd W
\,=\,
  S(t)\,W \,\dd t
\qqq
with the white noise $\,S(t)\,$ and $\,W(0) = \Id$.  \,The decoupling
of the Lagrangian trajectories $\,\bm R(t;\bm r_0))\,$ from these equations
is the great simplification of the Kraichnan model!
\vskip 0.1cm

The further simplification is the decoupling of the natural measures
from the statistics of $\,W(t;\bm v|\bm v))\,$ for the
homogeneous Kraichnan flows on a periodic box $\,V$. \,Due to the
homogeneity, such measures $\,n(\dd\bm r|\bm v)\,$ have to satisfy
the relation
$$
  \big\langle n(\dd\bm r|\bm v)\big\rangle
=
  \frac{\dd\bm r}{|V|}
$$
Now, since $n(\dd\bm r|\bm v)$ depends on the past velocities and, \,for $t
\geq 0$, $\,W(t,\bm r|\bm v)\,$ depends on the future ones (independent 
of the past ones in the Kraichnan ensemble), \,we infer that 
$$
  \big\langle \int_V f(W(t;\bm r|\bm v))\,\,n(\dd\bm r|\bm v)\big\rangle
=
  \int_V \big\langle f(W(t;\bm r|\bm v))\big\rangle\,
         \big\langle n(\dd\bm r|\bm v)\big\rangle
=
  \int_V \big\langle f(W(t;\bm r|\bm v))\,\frac{\dd\bm r}{|V|}\big\rangle
=
  \big\langle f(W(t;\bm r_0|\bm v))\big\rangle
\,.$$
It follows that the statistics of $\,W(t;\bm r|\bm v)\,$ for $\,t\geq0\,$ 
(and hence also of $\,\vec\rho(t;\bm r|\bm v)$\m) \,with $\,(\bm r,\bm v)\,$
distributed with the natural invariant measure $\,N(\dd\bm r,\dd\bm v)\,$ 
coincides with that of the solution of Eq.\,\eqref{eq:star2b}
with the white noise $\,S(t)\,$ and $\,W(0)=\Id\,$ (and of the corresponding
$\,\vec\rho(t)\,$). \,In other words,
$$P_t(\vec\sigma)\,=\,\tilde P_t(\vec\sigma)\qquad{\rm  for}\quad\  
t\geq0\,.$$ 
Since the Kraichnan ensemble is time reversible, we also infer 
from Eq.\,(\ref{TMFR}) that
$$
  P_t(-\antivec\sigma)
\,=\,
  P_t(\vec\sigma)\,\,\ee^{\m t\sum_{i=1}^d \sigma_i}
$$
for all $t\geq0$. 

\subsection{Multiplicative large deviations in Kraichnan velocities}
\subsubsection{Isotropic case}

The specific form of $\,P_t(\vec\sigma)\,$ depends
on the covariance $\,C^{ij}_{kl}\,$ reflecting the single-point
statistics of the velocity gradients. In the isotropic case, 
$$
  C^{ij}_{kl}
=
  \beta (\delta^i_k \delta^j_l + \delta^i_l \delta^j_k)
 +\gamma \delta^{ij} \delta_{kl}
$$
with the right hand side specified, up to normalization, by the
``compressibility degree''
\qq
  \wp
\,=\,
  \frac{\langle (\nabla_i v^i)^2 \rangle}
       {\langle (\nabla_j v^i)^2 \rangle}
\,=\,
  \frac{C^{ij}_{ij}}{C^{ii}_{jj}}
\,=\,
  \frac{(d+1)\beta + \gamma}{2\beta + \gamma d}\,.
\label{compd}
\qqq
In general, $\,0\leq\wp\leq1$. \,Vanishing $\,\wp\,$ corresponds 
to incompressible 
velocities, whereas $\,\wp=1\,$ to gradient ones.  The normalization
of $\,C^{ij}_{kl}\,$ is set by the time scale $\,\tau_\eta\,$ that we
shall define by the relation 
\qq
\tau_\eta\,=\,\frac{2d\,\delta(0)}{\langle(\nabla_i v^j)^2\rangle}\,=\, 
\frac{2d}{C^{ii}_{jj}}\,=\,\frac{2}{2\beta+\gamma d}\,.
\label{taue}
\qqq
It may be thought of as mimicking the Kolmogorov time scale in the
Kraichnan flow. By the It\^o calculus, for $\,W(t)\,$ solving the 
SDE (\ref{eq:star2b}), one obtains  the relation
\qq
 \frac{\dd}{\dd t}\big\langle f(W)\big\rangle
\,=\,
  \Big\langle \half W^m_i W^n_j C^{ij}_{kl}
    \frac{\partial}{\partial W^m_k} \frac{\partial}{\partial W^n_l}
  f(W)\Big\rangle\,.
\label{inav}
\qqq
Any function $\,\tilde f(\vec\rho)\,$ of the stretching exponents 
may be viewed as a function $\,f\,$ on the matrix group $\,GL(d)\,$ such 
that
$$
  f(\mathcal O' W \mathcal O)
\,=\,
  f(W)
\qquad\qquad
  \text{for\ \qquad$\mathcal O', \mathcal O \in O(d)\,$.}
$$
To calculate $\,\frac{\dd}{\dd t}\big\langle \tilde f(\vec\rho)\big\rangle$,
\,it is enough to compute the operator inside the average on the right hand
side of Eq.\,(\ref{inav}) in the action on such functions.
A straightforward although tedious calculation gives \cite{slow}: 
$$\frac{\dd}{\dd t}\,\big\langle \tilde f(\vec\rho)\big\rangle\,=\,
  \langle \mathscr L \tilde f(\vec\rho) \rangle$$ 
for the generator 
$$
  \mathscr L
\,=\,
  \frac{\beta+\gamma}{2}
  \Bigl(\sum_i \frac{\partial^2}{\partial\rho_i^2}
  +\sum_{i \neq j}
     \coth(\rho_i-\rho_j) \frac{\partial}{\partial\rho_i}
  \Bigr)
 +\frac{\beta}{2}
    \Bigl(\sum_i \frac{\partial}{\partial\rho_i}\Bigr)^2
 -\frac{(d+1)\beta+\gamma}{2} \sum_i \frac{\partial}{\partial\rho_i}
$$
After the conjugation by the function 
$\,\mathcal F(\vec\rho) = \ee^{-\frac12
  \sum_i\rho_i} (\prod_{i<j} \sinh(\rho_i-\rho_j))^{1/2}$,
\,the operator $\,\mathscr L\,$ becomes the Calogero-Sutherland-Moser 
Hamiltonian of $d$ quantum particles on the line \cite{OlshPer}: 
$$
  \mathcal F \mathscr L \mathcal F^{-1}
\,=\,
  \frac{\beta+\gamma}{2}
  \Bigl(\sum_i \frac{\partial^2}{\partial\rho_i^2}
  +\half \sum_{i<j}
     \frac{1}{\sinh^2(\rho_i-\rho_j)}
  \Bigr)
 +\frac{\beta}{2}
    \Bigl(\sum_i \frac{\partial}{\partial\rho_i}\Bigr)^2
 +\,\const
\ \equiv\ 
  -\m\mathscr H_{_{\rm CSM}}\,.
$$
Since, with the initial value $\,\vec\rho(0)=0$,
$$
  \big\langle \tilde f(\vec\rho(t))\big\rangle
\,=\,  \int \tilde f(t\vec\sigma)\,\,P_t(\vec\sigma)\,\,\dd\vec\sigma\,,
$$
it follows that
$$
  P_t(\vec\sigma)
\,=\,t\,\ee^{\m t\mathscr L}(\vec{0},t\vec\sigma)
\,=\,t\,
  \lim_{\vec\rho_0 \to 0}\,\,
    \mathcal F(\vec\rho_0)^{-1}
    \,\,\ee^{-t\mathscr H_{\rm CSM}}(\vec\rho_0,t\vec\sigma)
    \,\,\mathcal F(t\vec\sigma)\,.
$$
The spectral representation of the heat kernel $\,\ee^{-t\mathscr H_{\rm
CM}}\,$ is known explicitly \cite{OlshPer}.  Its saddle point 
calculation gives the large deviation form of 
$\,P_t(\vec\rho)$. \,The latter may be also found directly 
\cite{BalkFoux} since
$$
  P_t(\vec\sigma)\ 
\simeq\ 
  t\,\ee^{\m t\mathscr L_{\infty}}(0,t\vec\sigma)
\,=\,\const\,\,
  \ee^{-t H(\vec\sigma)}\,,
$$
where $\,\mathscr L_{\infty}\,$ is the second order 
differential operator with constant coefficients  
obtained from $\,\mathscr L\,$ by
replacing $\,\cosh(\rho_i-\rho_j)\,$ by $\,\pm 1\,$ for $\,i{<\atop>}j$,
\,respectively. \,An easy calculation gives then:
$$
  H(\vec\sigma)
\,=\,
  \frac{\tau_\eta}{4(d+\wp(d-2))}
  \Big[(d-1)(d+2)\sum_i (\sigma_i-\lambda_i)^2
   \,+\,(\wp^{-1}-d)
      \bigl(\sum_i (\sigma_i-\lambda_i) \bigr)^2\Big]
$$
with the Lyapunov exponents 
$$\,\lambda_i\tau_\eta\,=\,
\frac{(d+\wp(d-2))(d-2i+1)}{(d-1)(d+2)}\,-\,\wp
$$
equally spaced. In particular, $\,\lambda_1 > 0\,$ 
for $\,\wp < d/4\,$ (the chaotic 
phase) and $\,\lambda_1 < 0\,$ for $\wp > d/4$. \,The change of sign of
$\,\lambda_1\,$ induces a change in the transport properties (direct
cascade of the the scalar for $\,\lambda_1>0\,$ versus inverse cascade 
for $\,\lambda_1<0$ \cite{ChKolVerg}). The rate function $\,H\,$
satisfies the Gallavotti-Cohen relation (\ref{MFR}) with $\,H'=H$,
\,as may be easily checked. 
\vskip 0.1cm

The exponents $\,\gamma_q\,$
of Eq.\,(\ref{galp}) determining the asymptotic behavior of
the density fluctuations in the Lagrangian frame are
easily computed to be
$$
  \gamma_q\tau_\eta
\,=\,\wp\m d\,q(q+1)\,.
$$
Note that $\,\gamma_q> 0\,$ for $\,q> 0\,$ and $\,\wp>0\,$
so that in the compressible homogeneous isotropic Kraichnan flow 
the positive moments of the density grow exponentially in 
the Lagrangian frame.

\subsubsection{Non-isotropic case} 

The equal spacing of the Lyapunov exponents and the Gaussian form 
of the large deviations of the stretching rates
is particular for the Kraichnan flow that is homogeneous and
isotropic at short distances. On a periodic box, this is not
a generic situation. For example, on a periodic square, typically,
the covariance of the matrix-valued white noise $\,S(t)\,$ will
only have symmetries of the square and will be given by the expression
\qq
C^{ij}_{kl}\,=\,2\alpha\delta^{ij}_{kl}+\beta (\delta^i_k \delta^j_l 
+ \delta^i_l \delta^j_k) +\gamma \delta^{ij} \delta_{kl}\,,
\qqq
where $\,\delta^{ij}_{kl}\,$ is equal to 1 if $\,i=j=k=l\,$ and
vanishes otherwise. The compressibility degree (\ref{compd}) and
the Kolmogorov time (\ref{taue}) are now given by
$$
\wp\,=\,\frac{2\alpha+3\beta+\gamma}{2(\alpha+\beta+\gamma)}\,,
\qquad\tau_\eta\,=\,\frac{1}{\alpha+\beta+\gamma}
$$
and one may introduce the ``anisotropy degree''
$$\\omega\,=\,|\alpha|\tau_\eta\,.$$
The Lyapunov exponents may be expressed in this case \cite{CDG} by
the complete elliptic integrals \cite{GrRh}:
\qq
{\lambda_1\tau_\eta}\, 
=\,-\,1\,+\,{3-2\wp+\omega\over 2}\,{\bm E(k)
\over\bm K(k)}\,,\qquad\quad
{\lambda_2\tau_\eta}\,=\, 
1\,-\,2\wp\,-\,{3-2\wp+\omega\over 2}\,{\bm E(k)
\over\bm K(k)}\,,
\nonumber
\qqq
where $\,k^2=\frac{2\omega}{3-2\wp+\omega}$, \,depicted 
below as functions of $\,0\leq\omega\leq 1\,$
for three values of the compressibility degree 
$\,\wp=0$, $\,\wp=\frac{1}{2}\,$ and $\,\wp=1\,$ (the middle green lines give
$\,\frac{1}{2}(\lambda_1+\lambda_2)\tau_\eta=-\wp$).

\begin{figure}[ht]
\vskip -0.2cm
  \begin{center}
    \subfigure[\label{fig:p0} $\wp = 0$]{%
      \begin{minipage}{0.24\textwidth}
        \includegraphics[width=\textwidth]{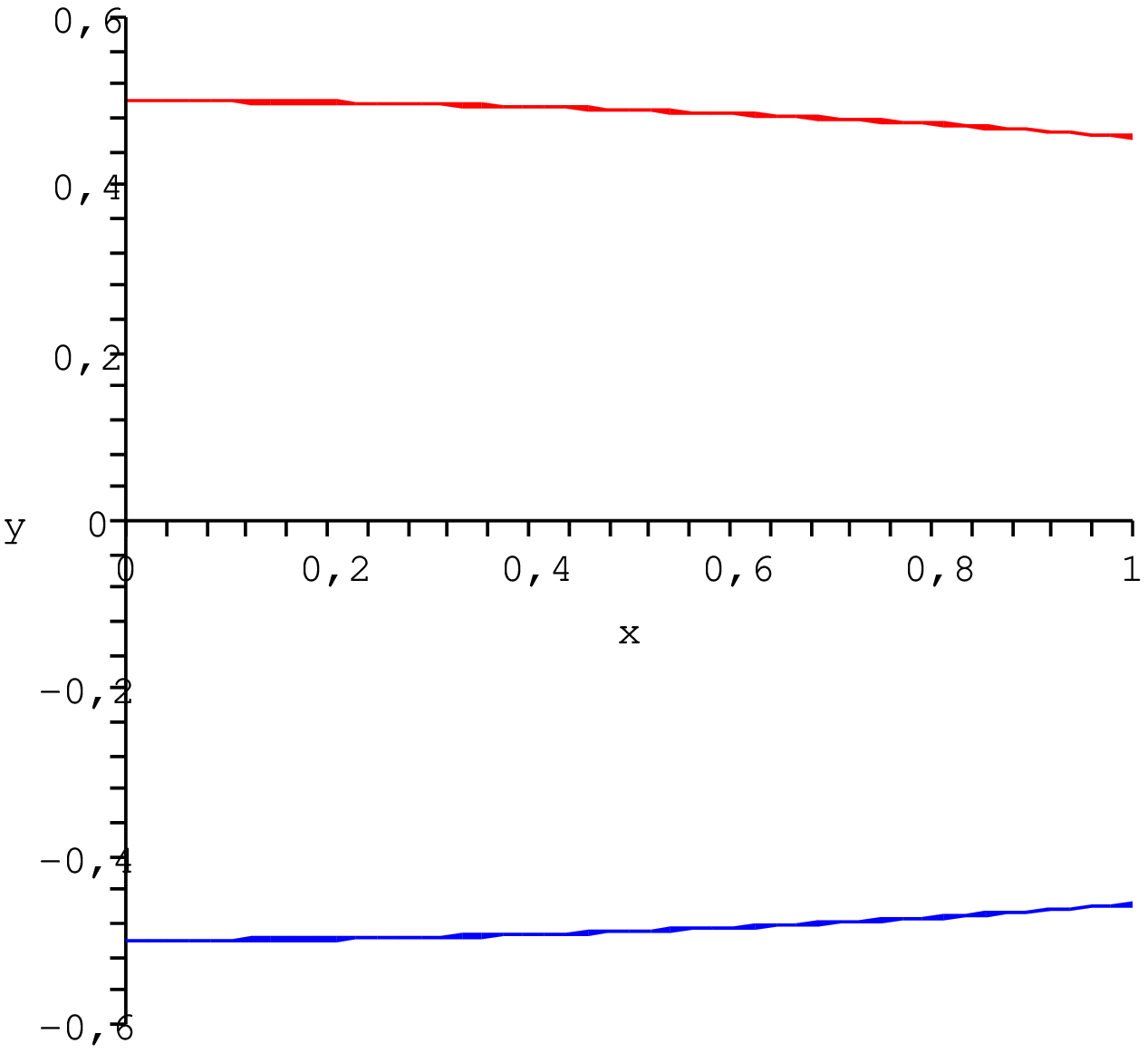}\\
        \vspace{-0.8cm} \strut
        \end{minipage}}
    \hspace{0.5cm}
    \subfigure[\label{fig:p0.5} $\wp = {1\over2}$]{%
      \begin{minipage}{0.24\textwidth}
        \includegraphics[width=\textwidth]{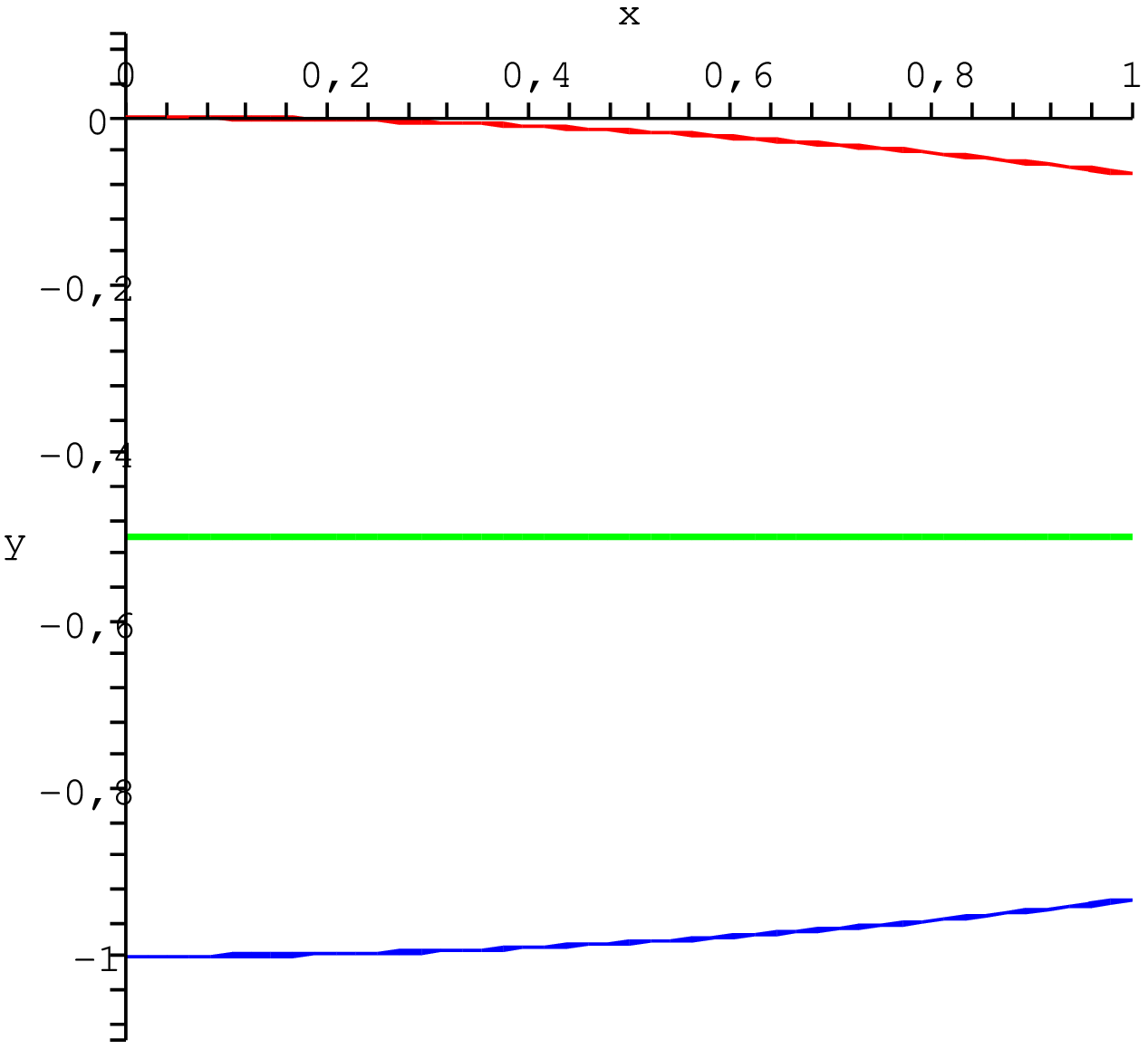}\\
      \vspace{-0.8cm} \strut
        \end{minipage}}
    \hspace{0.5cm}
    \subfigure[\label{fig:p1} $\wp = 1$]{%
      \begin{minipage}{0.25\textwidth}
      \vspace{0.3cm} 
       \includegraphics[width=\textwidth,height=3.5cm]{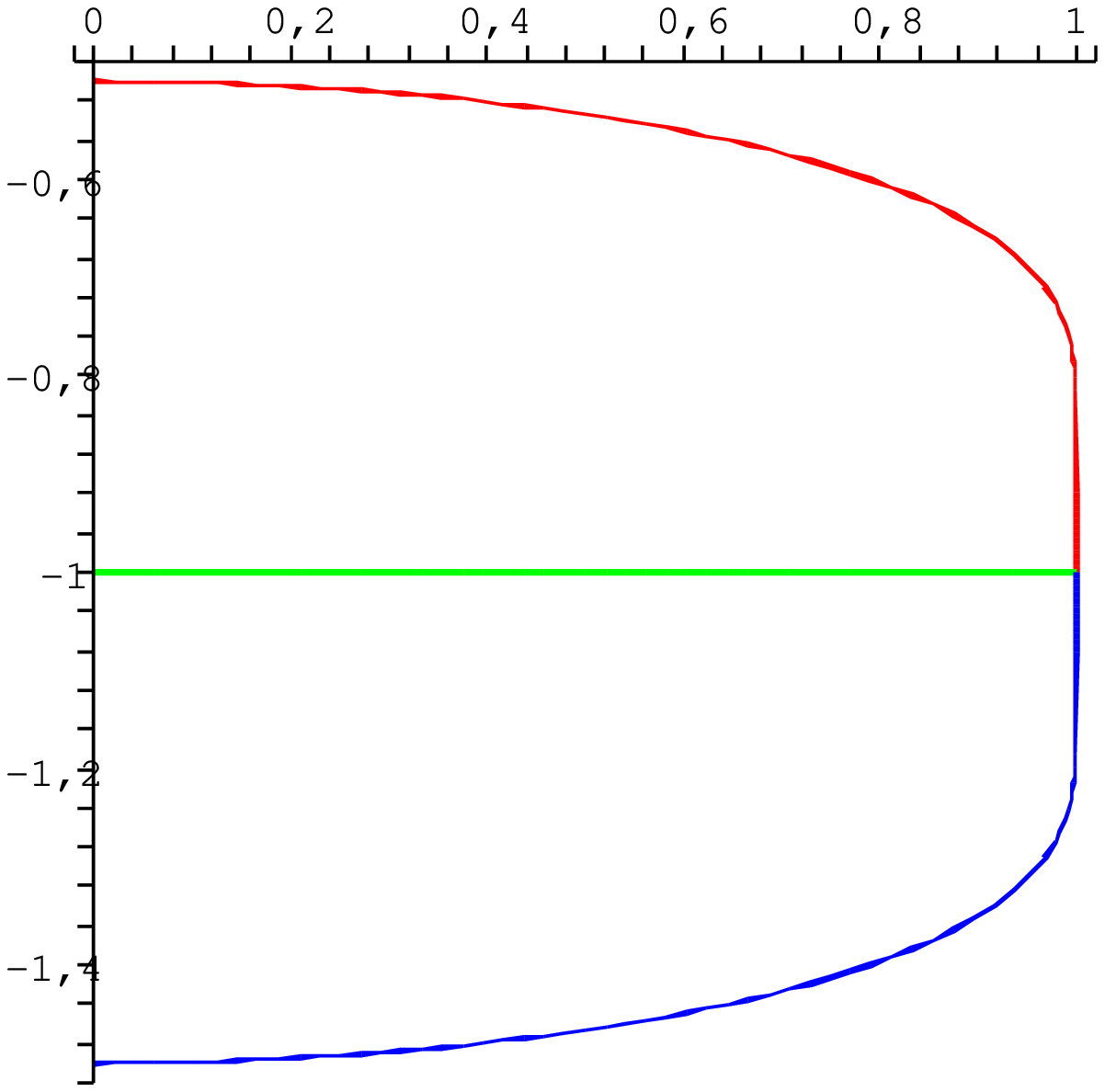}\\
      \vspace{-0.45cm} \strut
        \end{minipage}}
    \vspace{-0.4cm}
\end{center}
\vspace{0.2cm}
\end{figure}
  
\noindent If $\,\lambda_1>\lambda_2\,$ then the large deviations 
of the stretching exponents are controlled by the rate function\cite{CDG}
\qq
H(\sigma_1,\sigma_2)&=&\frac{_1}{^8}\tau_\eta\wp^{-1}(\sigma_1+\sigma_2
+2\wp\tau_\eta^{-1})^2\cr\cr
&+&\max\limits_\nu\Big[\nu(\sigma_1-\sigma_2)
-(1+\omega)\tau_\eta^{-1}\nu(\nu+1)+(3-2\wp+\omega)\tau_\eta^{-1}\,
E_{\nu,0}(k^2)\Big], 
\label{LDRF}
\qqq
where $\,E_{\nu,0}(k^2)\,$ is the groundstate energy of the Lam\'e-Hermite
one-dimensional Schr\"odinger operator
\qq
-\frac{{\dd^2}}{{\dd u^2}}\,+\,\nu(\nu+1)\,k^2\,{\rm sn}^2(u,k)
\qqq
acting on periodic functions of $\,u\,$ with period $\,2\bm K(k)$.
$\,{\rm sn}(u,k)\,$ is the Jacobian sine-amplitude function
\cite{GrRh}. \,If $\,\lambda_1=\lambda_2$, \,which happens 
for $\,\omega=1=\wp$, \,then
\qq
H(\sigma_1,\sigma_2)\,&=&\,\frac{_1}{^8}\tau_\eta(\sigma_1+\sigma_2
+2\tau_\eta^{-1})^2\,+\,\frac{_1}{^8}\tau_\eta(\sigma_1-\sigma_2)^2
\qqq
and cannot be obtained as the limit of the previous formula (the time-scales
at which the large deviation regime sets in diverge as 
$\,(\lambda_1-\lambda_2)\tau_\eta\to0\,$ \cite{CDG}). \,Except for the last 
case, the large deviations of $\,(\rho_1-\rho_2)\,$ are non-Gaussian 
in the presence of anisotropy, i.e.\ for $\,\omega>0$.

\subsection{Inertial particles in Kraichnan model}

The inertial particles dynamics in the Kraichnan velocities is given
for the vanishing diffusivity $\kappa$ by the SDE
\qq
\dd\bm R\,=\,\bm U\,\dd t\,,\qquad\dd\bm U\,=\,\frac{_1}{^\tau}\big(-\bm U 
+ \bm v(t,\bm R)\big)\,\dd t\,,   
\label{IPK}
\qqq 
see Eq.\,(\ref{IP}). Both It\^o or Stratonovich conventions give for it
the same result. Small phase-space separations between two inertial particles
satisfy, in turn, the equations
\qq
\dd\bm\delta R\,=\,\bm \delta V\,\dd t\,,\qquad\dd\delta\bm U\,
=\,\frac{_1}{^\tau}\big(-\delta\bm U+\Sigma(t)\m\delta\bm R\big)\,\dd t\,,
\label{dRVK}
\qqq 
where, as before, $\,\Sigma^i_{\,\,j}(t)=\nabla_jv^i(t,\bm R(t))$.
The matrix $\,\bm W(t;\bm r,\bm u)\,$ propagating 
$\,(\delta\bm R,\delta\bm U)\,$ along the trajectory starting at 
$\,(\bm r,\bm u)\,$ satisfies then the SDE
\qq
d\bm W\,=\,\Big(\begin{matrix}0&\Id\cr\frac{1}{\tau}\Sigma(t)&-\frac{1}{\tau}
\end{matrix}\Big)\,\bm W\,\dd t\,.
\label{bmWK}
\qqq
As before, if the velocity statistics is homogeneous then $\,\Sigma(t)\,$ 
may be replaced by the matrix-valued white noise $\,S(t)\,$ as long as we 
are interested in the statistics of $\,\bm W(t;\bm r,\bm u)\,$ for fixed 
$\,(\bm r,\bm u)$. \,The same is true if we average over the latter points 
with the natural measures $\,n(\dd\bm r,\dd\bm u)\,$ since the 
latter average decouples due to the independence of the past and future 
velocities). For the homogeneous and isotropic Kraichnan 
velocities, on a periodic box $\,V$, \,the ensemble average of the 
natural measures is equal to the Maxwell distribution:
\qq
\big\langle n(\dd\bm r,\dd\bm u)\big\rangle\,=\,\frac{_{\tau^{d/2}}}
{^{(2\pi D_0)^{d/2}|V|}}\,\ee^{-\frac{\tau}{2D_0}\bm U^2}\,\dd\bm r\,
\dd\bm u\,.
\nonumber
\qqq
Note that Eq.\,(\ref{bmWK}) implies that $\,\det\bm W(t)=\ee^{-d\m t/
\tau}\,$ so that the corresponding stretching rates 
$\,\sigma_i(t),\ i=1,\dots,2d,$ \,satisfy the relation
\qq 
\sum\limits_{i=1}^{2d}\sigma_i(t)\,=\,-\frac{d}{\tau}\,.
\label{sripK}
\qqq
A simple argument \cite{FouxHorv} shows that the Gallavotti-Cohen
type symmetry (\ref{MFR}) for the large deviations rate functions 
reduces now to the identity
\qq
H(\tau^{-1}\vec{1}-\antivec{\sigma})\,=\,H(\vec\sigma)
\qqq
with $\,\vec{1}\equiv(1,\dots,1)\,$ holding for $\,\vec{\sigma}\,$ 
satisfying the relation \,(\ref{sripK}).

\subsubsection{One-dimensional case}

For the one-dimensional Kraichnan velocities, setting 
$\,\delta r(t)=\psi(t)\,\ee^{\m t/(2\tau)}\,$ permits to rewrite 
Eqs.\,(\ref{dRVK}) in the second order form \cite{WM}
\qq
-\ddot\psi\,+\,V(t)\m\psi\,=\,-\frac{1}{4\tau^2}\psi
\label{schr}
\qqq
with $\,V(t)=\frac{1}{\tau}\Sigma(t)$. \,Again, $\,V(t)\,$ may be replaced 
by a white noise with the covariance 
\qq
\big\langle V(t)\,V(t')\,=\,\frac{2}{\tau^2\tau_\eta}\m\delta(t-t') 
\qqq
Upon viewing $\,t\,$ as a space variable, Eq.\,(\ref{schr}) becomes
the one-dimensional stationary Schr\"odinger equation in the delta-correlated
potential $\,V(t)\,$ at the energy $\,E=-\frac{1}{4\tau^2}$, a well
known model for the one-dimensional Anderson localization. 
This is an old problem whose solution goes back to \cite{Halp}, see
also \cite{LGP}. One considers the process $\,x(t)=\dot\psi(t)/\psi(t)\,$
that satisfies the SDE
\qq 
\dd x\,=\,-(x^2+E-V(t))\,\dd t\,.
\qqq
Its trajectories reach $\,-\infty\,$ in a finite time but reappear 
immediately at $\,+\infty\,$. \,Such jumps correspond to a passage
of $\,\psi(t)\,$ through zero with $\,\dot\psi\not=0\,$ or to
a crossing of close inertial particles in space. The jumping process
possesses the stationary probability measure 
\qq
n(\dd x)\,=\,\frac{1}{Z}\,\ee^{-\tau^2\tau_\eta(\frac{1}{3}x^2+Ex)}
\Big(\int\limits_{-\infty}^x\ee^{\tau^2\tau_\eta(\frac{1}{3}y^2+Ey)}\dd y\Big)
\,\dd x\,,
\qqq
with the density $\,\sim\,x^{-2}\,$ for large $\,|x|$ \,($\,Z\,$ 
is the normalization constant). The localization
Lyapunov exponent $\,\lambda$, \,equal to the mean value of 
$\,x(t)=\frac{\dd}{\dd t}\ln(|\psi(t)|)$, \,is always positive
signaling the permanent localization in one dimension. 
The Lyapunov exponents for the inertial particles are 
$\,\lambda_1=\lambda-\frac{1}{2\tau}\,$ and $\,\lambda_2=-\lambda-
\frac{1}{2\tau}\,$ (recall the relation between
$\,\psi(t)\,$ and $\,\delta r(t)$). $\,\lambda\,$ can be expressed \cite{LGP}
by the Airy functions \cite{GrRh} and one obtains:
\qq 
\lambda_{1,2}\tau_\eta\ =\ 4u^2\Big(-u\,\pm\,\frac{\Ai'(u^2)
\,\Ai(u^2)\,+\,\Bi'(u^2)\,\Bi(u^2)}{\Ai(u^2)\,+\,\Bi(u^2)}\Big)\,,
\qqq
for $\,u=\frac{1}{2}\StNo^{-1/3}\,$ and the Stokes number
$\,\StNo=\frac{\tau}{\tau_\eta}$. \,For large $\,\StNo$, \,$\lambda_1\,$
is positive and decreases as $\,\StNo^{-2/3}$. \,For $\,\StNo\,$ small,
it is negative and approaches when $\,\StNo\to0\,$ the value $\,-1\,$
for the Lagrangian particles.

\subsubsection{Two-dimensional case}

For the two-dimensional homogeneous and isotropic Kraichnan velocities,
one may treat $\,\delta\bm r(t)\,\ee^{t/(2\tau)}=\psi(t)\,$ as a 
complex-valued process which still satisfies Eq.\,(\ref{schr})
with $\,V(t)\,$ a complex valued white noise with the covariance
\cite{Piterb,MWDWL}
\qq
\big\langle V(t)\,V(t')\big\rangle\,
=\,\frac{2\wp-1}{\tau^2\tau_\eta}\m\delta(t-t')\,,
\qquad
\big\langle V(t)\,\overline{V(t')}\big\rangle\,=\,\frac{2}{\tau^2\tau_\eta}
\m\delta(t-t')\,. 
\qqq
For the complex-valued process $\,z(t)=\dot\psi(t)/\psi(t)\,$ one obtains
now the SDE
\qq
\dd z\,=\,-(z^2+E-V(t))\,\dd t
\qqq
which may be shown to possess an invariant probability measure, this time
without an explicit analytic form but decaying like $\,|z|^{-4}\,$ for
large $\,|z|$. \,The top Lyapunov exponent $\,\lambda_1\,$
is equal to the expectation of $\,{\mathrm Re}\,z\,$ in this measure. 
It was studied numerically in \cite{MWDWL,Horv,BCH0}. It seems again
to behave as $\,\StNo^{-2/3}\,$ for large Stokes numbers. The reference
\cite{BCH0} contains also the numerical results about large
deviations of $\,{\mathrm Re}\,z(t)$.

\begin{conc}
In smooth Kraichnan velocities with symmetry properties, the tangent
process for Lagrangian particles is related to integrable quantum-mechanical 
systems. This relation gives an analytic access to the large deviation 
rate functions of the stretching exponents. For inertial particles,
similar relations to one-dimensional models of the Anderson localization
permit to find analytically the Lyapunov exponents in one-dimensional flows 
and facilitate their analysis in two (and three) dimensions.
\end{conc}

\newpage

\nsection{Particles and fields in rough Kraichnan velocities}

The previous discussion pertained to the transport of particles
in the Batchelor regime with smooth velocities and small particle
separations. We shall pass now to the study of particle dynamics
at separations in the inertial range of scales where velocities
become rough. We shall concentrate on the motion of Lagrangian
particles. For recent ideas and numerics about the motion of inertial
particles in rough flows, see \cite{BCH,FH}. 
\vskip 0.1cm

Let us start by considering 
the simultaneous motion of $N$ Lagrangian particles $\,\bm R_n(t)\,$
in the homogeneous Kraichnan velocities.  
A function $f(\underline{\bm R})$ of their
joint positions $\underline{\bm R}(t) \equiv (\bm R_1(t),\dotsc,\bm R_N(t))$
evolves accordingly to the Ito stochastic equation
\qq
  \dd f(\underline{\bm R})
\ =\ 
\sum_{n=1}^N v^i(t,\bm R_i)\,\nabla_{R_n^i} f(\underline{R}) \,\dd t
\ +\ \underbrace{
    \frac12 \sum_{m,n=1}^N D^{ij} (\bm R_m - \bm R_n)
    \nabla_{R_m^i} \nabla_{R_m^j}}_{\mathcal M_{\hspace{-0.03cm}N}}
  f(\underline{R}) \,\dd t\,.
\qqq
For the expectations, this gives:
\qq
  \frac{\dd}{\dd t}\big\langle f(\underline{\bm R})\big\rangle
\,=\,
  \big\langle (\mathcal M_{\hspace{-0.03cm}N} f)
(\underline{\bm R})\big\rangle\,.
\label{MN}
\qqq

\subsection{Particle dispersion}

First, let us study the evolution of a function of the separation
$\,\Delta \bm R(t) \equiv \bm R_1(t) - \bm R_2(t)\,$ 
between two trajectories: 
\qq
  \frac{\dd}{\dd t}\big\langle f(\Delta\bm R)\big\rangle
\,=\,
  \big\langle
    (D^{ij}(\bm 0) - D^{ij}(\Delta \bm R ))
    \nabla_i \nabla_j f(\Delta \bm R)
  \big\rangle\,.
\label{evsep}
\qqq
In particular, we shall look at functions of the inter-particle distance
$\,\Delta\equiv |\Delta \bm R|$, \,called the ``particle dispersion''.
\,In the Batchelor regime,
$$
  D^{ij}(\bm 0) - D^{ij}(\Delta \bm R )
\ \approx\ 
  \frac12 \Delta R^k \Delta R^l\,\nabla_k \nabla_l D^{ij}(\bm 0)
$$
and Eq.\,(\ref{evsep}) reduces to the same equation as for 
$\,\big\langle f(\delta \bm R) \big\rangle$, \,see the relation 
(\ref{fita}).  \,For functions of the particle dispersion $\,\Delta
\equiv |\Delta \bm R|$, \,one obtains after a short calculation,
assuming also isotropy of the velocity ensemble:
$$
  \frac{\dd}{\dd t}\big\langle f(\Delta)\big\rangle
\,=\,
  \Big\langle \Big[
    \frac{2\beta+\gamma}{2}
    \Big(\frac{\partial}{\partial\ln\Delta} \Big)^2
   +\lambda_1 \frac{\partial}{\partial\ln\Delta}
  \Big] f(\Delta) \Big\rangle\m,
$$
where $\,\lambda_1 = \frac{d-2}{2}\gamma - \beta\,$ is the first Lyapunov
exponent.  For the PDF of the distance $\,\Delta\,$ this gives
$$
  \big\langle\delta(\Delta-|\Delta \bm R(t)|)\big\rangle
\,=\,
  \frac{1}{\sqrt{2\pi(\beta+\gamma)t}}\,
  \ee^{-\frac{(\ln\frac{\Delta}{\Delta_0}-\lambda_1 t)^2}{2(\beta+\gamma)t}}
  \,\frac{1}{\Delta}\,,
$$
i.e.\ a log-normal distribution with $\,\Delta_0 \equiv |\Delta \bm
R(0)|$.  \,Note that
$$
  \lim_{\Delta_0 \to 0}
    \,\,\big\langle \delta(\Delta - |\Delta R(t)|)\big\rangle
=
  \delta(\Delta)
$$
so that when the initial distance $\,\Delta_0\,$ of the trajectories tends 
to zero, so does their time $\,t\,$ distance. The latter behavior is
imposed by the uniqueness of the Lagrangian trajectories in each velocity 
realization, given their initial point (and assuming that the molecular
diffusion is absent). \,It is consistent with both
exponential separation (when $\lambda_1>0$) and exponential contraction
(when $\lambda_1<0$) of close trajectories.
\vskip 0.1cm

Now, let us consider the Kraichnan model in the inertial range.  Here
$\,D^{ij}(\bm 0) - D^{ij}(\bm r) = O(r^\xi)\,$ and the isotropy imposes 
the form of the tensor:
$$
  D^{ij}(\bm 0) - D^{ij}(\bm r)
\,=\,
  \beta\, r^i r^j |\bm r|^{\xi - 2} + \frac12 \gamma\, \delta^{ij} r^\xi
$$
in the way fixed, up to normalization, by the compressibility
degree
$$
  \wp
\,=\,
  \lim_{r\to0} \frac{\nabla_i \nabla_j D^{ij}(\bm r)}
                    {\nabla_j \nabla_j D^{ii}(\bm r)}
\,=\,
  \frac{2\xi^{-1}(d-1+\xi)\beta + \gamma}{2\beta +d\gamma}
$$
between 0 and 1 \,(taking $\,\xi=2\,$ reproduces the smooth case formula
(\ref{compd})). \,The equation for the evolution of the two-particle dispersion
$\,\Delta\,$ takes now the form
$$
  \frac{\dd}{\dd t}\big\langle f(\Delta)\big\rangle
\,=\,
  \Big\langle
    \frac{2\beta+\gamma}{2} \Delta^{\xi-a}
    \frac{\partial}{\partial\Delta}\Delta^a\frac{\partial}{\partial\Delta}
    f(\Delta)
  \Big\rangle\ \equiv\ \big\langle(M_a f)(\Delta)\big\rangle
$$
for $\,a=\frac{(d-1)\gamma}{2\beta+\gamma}=\frac{d+\xi}{1+\wp\xi}-1$.
\,The distance process $\,\Delta(t)\,$ is the one-dimensional diffusion 
with the generator $\,M_a\,$ and the transition PDF
$$
  \big\langle\delta(\Delta - \Delta(t;\Delta_0))\big\rangle
\,=\,
  \ee^{\m tM_a}(\Delta_0,\Delta)\,.
$$
It is instructive to change variables from $\,\Delta\,$ to $\,x=
\Delta^{1-\xi/2}\,$ since the process $\,x(t)\,$ 
is the Bessel diffusion with the generator proportional to
$\,x^{1-D_\eff} \partial_x x^{D_\eff-1}\partial_x$, \,i.e.\ to the radial 
Laplacian in $D_\eff$ dimensions, where
$$
  D_\eff
\,=\,
  \frac{2(a+1-\xi)}{2-\xi}
$$
In other words, $\,x(t)\,$ behaves as the norm of the Brownian motion in
(continuous) dimension $\,D_\eff$. \,The short-distance behavior 
of the Brownian motion below and above two dimensions is different. 
In particular, for $\,0<D_\eff<2$, \,i.e.\ for
$$
  p_c^1
\,\equiv\,
  \frac{d-2}{2\xi}+\frac12
\,<\,
  \wp
\,<\,
  \frac{d}{\xi^2}
\,\equiv\,
  p_c^2\,,
$$
the generator of the Bessel process admits different boundary conditions 
at $\,x=0$. \,They correspond to different behaviors of Lagrangian 
particles when they collide.

\subsection{Phases of Lagrangian flow}

There are three different phases with trajectory behavior and different 
transport properties in the inertial range, depending on the level 
of compressibility \cite{GV,EV1,LeJRai1}.

\subsubsection{Weakly compressible phase}

For $0 \leq \wp
  \leq p_c^1$, \,i.e. for $\,D_\eff>2$, \,there is only one possible boundary
  condition at zero:
$$
  (\Delta^a \frac{\partial}{\partial\Delta}f)(0)
\,=\,0
$$
and it renders $\,M_a\,$ self-adjoint on the Hilbert space
$L^2(\RR_+,\Delta^{a-\xi} \dd\Delta)$.  \,The process $\,\Delta(t)$ has 
a reflecting behavior at zero and
$$
  \lim_{\Delta_0 \to 0}
    \langle \delta(\Delta-\Delta(t;\Delta_0)) \rangle
\,=\,
  \const\,\, \frac{1}{t^{\frac{a+1-\xi}{2-\xi}}}
  \,\,\ee^{-\frac{2\Delta^{2-\xi}}{(2\beta+\gamma)(2-\xi)^2t}}\m.
$$
The formula says that the distance between two particles at time $t$ 
is smoothly distributed even if their initial distance is zero: the 
trajectories separate in finite time (unlike in the chaotic regime 
with exponential separation where close particles take longer and
longer to attain a sizable separation)! \,This implies the 
``spontaneous randomness'' of the Lagrangian flow \cite{slow}: \,the 
sharp distribution  
$\,\delta(\bm r-\bm R(t;t_0,\bm r_0|\bm v)\,$ of the time $\,t\,$ position 
of the trajectory starting at a fixed point in a fixed velocity field
realization is replaced by a diffused distribution $\,P(t_0,\bm r_0;t,
\bm r|\bm v)\,$ that is not concentrated at one point, as illustrated below:

\begin{figure}[h!]
\begin{center}
\vspace{0.1cm}
\mbox{\hspace{0.0cm}\epsfig{file=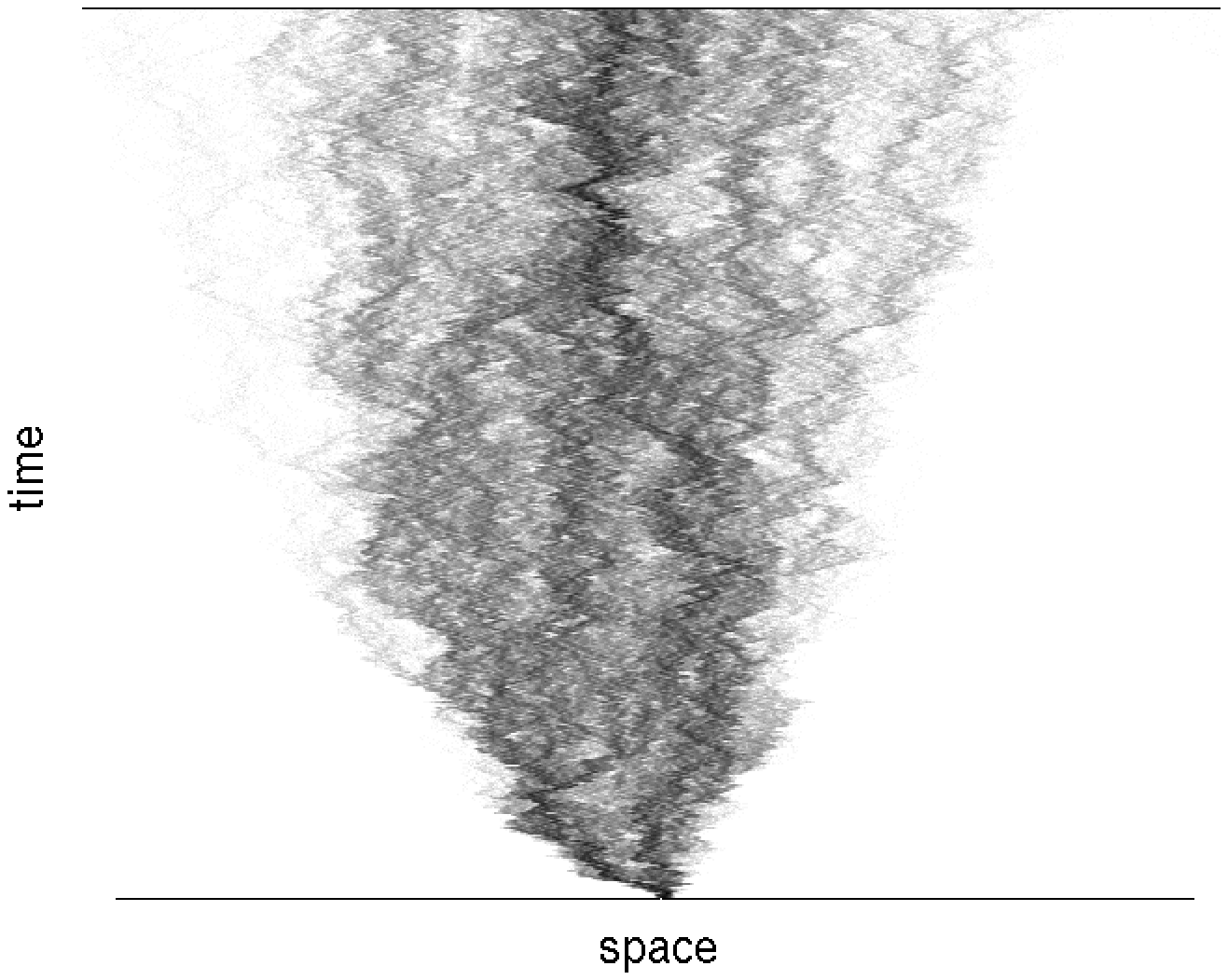,height=4.5cm,width=5cm}}
\end{center}
\vskip -0.1cm
\end{figure}

\noindent This is possible because for $\,D(\bm 0)-D(\bm r) \propto 
|r|^\xi\,$ the typical velocity realizations $\bm v(t,\bm r)$
are only H\"older continuous in space, with exponent smaller than $\,\xi/2$. 
In such velocity fields, the solutions of the trajectory equation 
$\,\frac{\dd \bm R}{\dd t} = \bm v(t,\bm R)\,$ 
in a fixed $\,\bm v\,$ are not uniquely determined by the initial condition 
but form a ``generalized flow'': a
stochastic (Markov) process with transition probabilities $\,P(t_0,\bm
r_0;t,\bm r|\bm v)$. \,The latter have been constructed rigorously by 
Le~Jan and Raimond in \cite{LeJRai1}.
\vskip 0.1cm

The spontaneous randomness of the Lagrangian flow implies the persistence 
of dissipation in the scalar transport. Note that for the incompressible 
flow where
$$
  \int P(t,\bm r;t_0,\bm r_0|\bm v)\,\,\dd\bm r_0 
\,=\,
  \int P(t,\bm r;t_0,\bm r_0|\bm v)\,\,\dd\bm r
\,=\,
  1
$$
and the unforced scalar evolves according to the equation
\qq
\theta(t,\bm r)\,=\,\int P(t,\rm r;t_0,\bm r_0)\,\,\theta(t_0,\bm r_0)\,\,
\dd\bm r_0\,,
\qqq
the inequality
\begin{align*}
  0
\leq
  \int P(t,\bm r;t_0,\bm r_0|\bm v)
    \,\,(\theta(t_0,\bm r_0)-\theta(t,\bm r))^2\,\dd \bm r_0\,\dd \bm r
\,=\,
  \int \theta(t_0,\bm r_0)^2 \,\dd r_0
 -\int \theta(t  ,\bm r  )^2 \,\dd r
\end{align*}
implies that the $L^2$ norm of $\,\theta\,$ (the scalar ``energy'') cannot
grow. Besides, it is conserved if and only if, for each $\,\bm r$,
$\,\theta(t_0,\bm r_0) = \theta(t,\bm r)\,$ on the support of $\,P(t,\bm
r;t_0,\cdot|\bm v)$, so if and only if the flow is deterministic.
The persistent dissipation of the scalar energy in generalized flows
is a short-distance phenomenon. It leads to the direct scalar-energy 
cascade where the scalar energy supplied at long distances is transfered 
without loss towards shorter and shorter scales and finally
dissipated on the infinitesimal ones (in the limit of vanishing
molecular diffusivity $\kappa$) \cite{Kr68,FGV}.

\subsubsection{Strongly compressible phase}

For $\,\wp > p_c^2$, \,i.e.\ for $\,D_\eff < 0$, \,the only possible 
boundary condition for  $\,M_a\,$ is $\,f(0)=0\,$ and
it again renders $\,M_a\,$ self-adjoint in
$\,L^2(\RR_+,\Delta^{a-\xi}\dd\Delta)$. \,Now, $\,\Delta(t)$ is absorbed 
at zero and
$$
  \big\langle \delta(\Delta-\Delta(t;\Delta_0))\big\rangle
\ =\ 
  \text{regular}\,+\,\const\,\delta(\Delta)\,,
$$
see \cite{GV}. In this phase there is a positive probability for the Lagrangian
trajectories that start at positive distance $\,\Delta_0$ \,to collapse 
together by time $\,t$.  \,When $\,\Delta_0 \to 0\,$ then the contact 
term dominates:
$$
  \lim_{\Delta_0 \to 0}
    \,\,\big\langle \delta(\Delta-\Delta(t,\Delta_0))\big\rangle
\,=\,
  \delta(\Delta)
$$
which signals that the trajectories are deterministic:
$$
  P(t_0,\bm r_0;t,\bm r|\bm v)
\,=\,
  \delta(\bm r - \bm R(t;t_0,\bm r_0|\bm v))
$$\begin{figure}[h!]
\begin{center}
\vskip -0.45cm
\leavevmode
{%
      \begin{minipage}{0.19\textwidth}
        \includegraphics[width=1.3\textwidth]{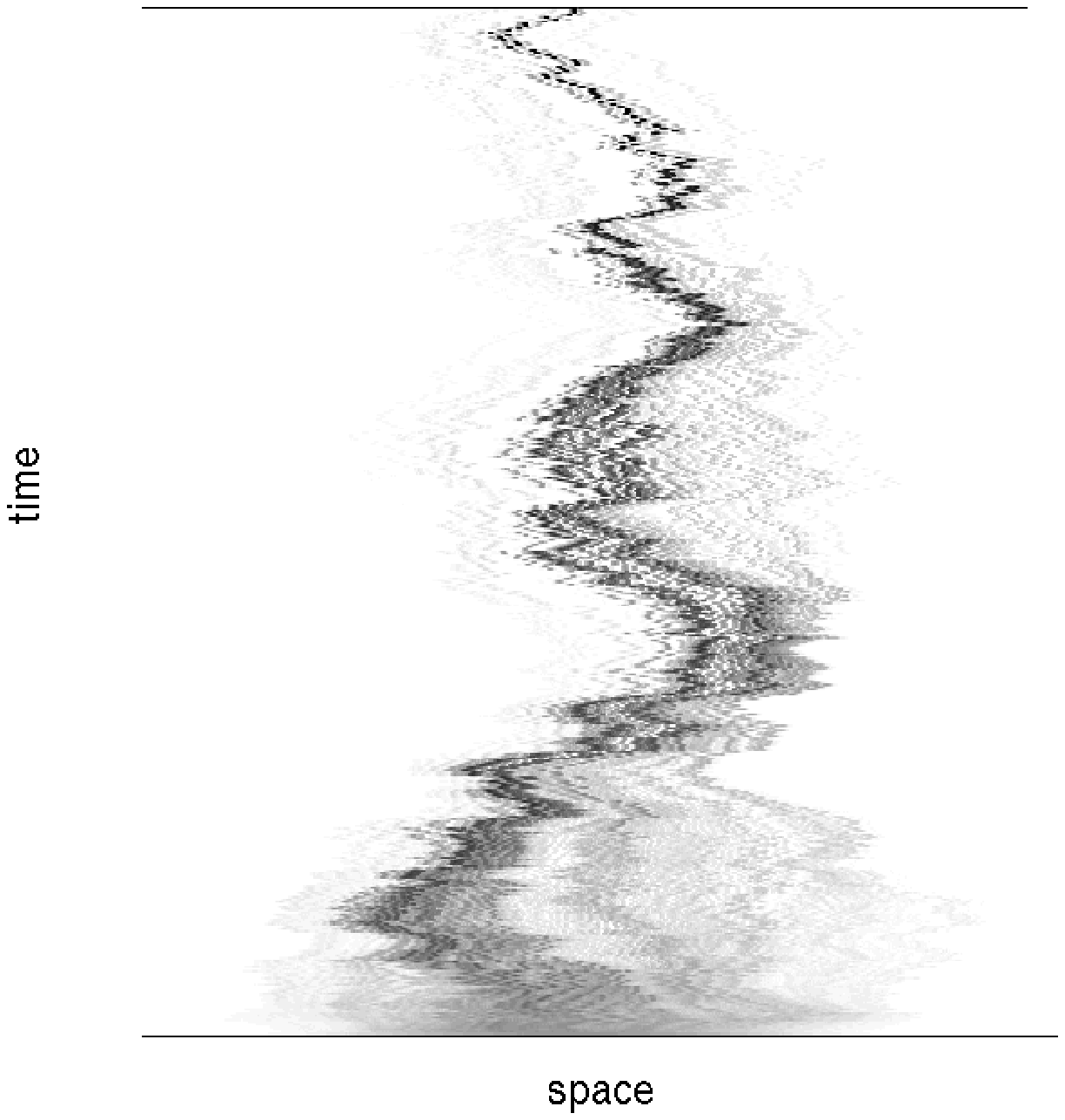}\\
        \vspace{-0.8cm} \strut
        \end{minipage}}
    \hspace{1.6cm}
{%
      \begin{minipage}{0.255\textwidth}\vspace{0.1cm}
        \includegraphics[width=\textwidth]{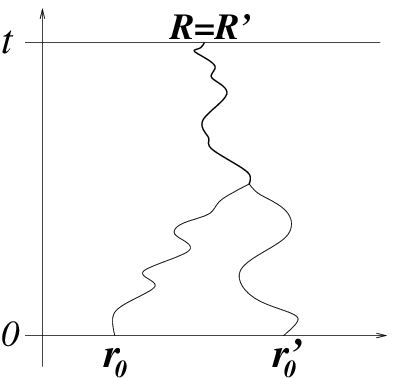}\\
      \vspace{-1.5cm} \strut
        \end{minipage}}\hspace*{20pt}
\end{center}
\end{figure}
for single trajectories $\,\bm R(t;t_0,\bm r_0|\bm v)\,$ collapsing for
different $\,\bm r_0\,$ at later times, as on the illustration:
\vskip -0.15cm


\noindent This is again a non-standard behavior.
Since trajectories are unique, there is no persistent dissipation of 
the scalar energy and, in the presence of a steady source, the latter
exhibits an inverse cascade (towards long distances) \cite{GV}.

\subsubsection{Intermediate compressibility phase}

For $\,p_c^1< \wp < p_c^2$, \,i.e.\ $\,0<D_\eff<2$, \,both reflecting 
and absorbing boundary conditions are possible, the first one is selected 
by adding very small diffusivity, the second one by considering velocities 
smeared at very small distances (small viscosity) \cite{EV1}.
The other permitted boundary conditions are the ``sticky''
ones \cite{GawHorv}:
$$
  \mu\,\Big(\Delta^{\xi-a}\frac{\partial}{\partial\Delta}\Delta^a
\frac{\partial}{\partial\Delta}f\Big)(0)
\,=\,
  \Big(\Delta^a\frac{\partial}{\partial\Delta}f\Big)(0)
$$
for $\,\mu$ the amount of ``glue'', $\,0 < \mu < \infty$.  \,These
conditions render $\,M_a\,$ self-adjoint on the space
$\,L^2(\RR_+,(\Delta^{a-\xi} + \mu\m\delta(\Delta))\dd\Delta)$.  The
particles stick together and separate immediately, never spending a
finite interval of time together. Nevertheless, they spend a finite
portion of the total time glued together.  The persistent dissipation 
is still present and leads to a direct scalar-energy cascade of the forced scalar.  
It induces however an anomalous scaling of the 2-point scalar structure
function $\,\big\langle(\theta(t,\bm r)-\theta(t,\bm 0))^2\big\rangle 
\propto r^{1-a}\,$ (instead of the dimensional result $\,\propto r^{2-\xi}$, 
\,see below). The sticky condition corresponds to fine-tuned small 
diffusivity and small viscosity \cite{EV2,GawHorv}.
\vskip 0.1cm

For $\,p_c^1 < \wp < p_c^2$, \,the transition probabilities 
$\,P(t_0,\bm r_0;t,\bm r|\bm v)\,$ were constructed rigorously by Le~Jan and
Raimond\cite{LeJRai1,LeJRai2}, for the reflecting boundary
condition, the absorbing one and, in one dimension, also for the sticky ones.
In the first case, they depend on $\bm v$ only, whereas in the second 
and third cases, on the (white noise) $\,\bm v\,$ and on the additional 
``black'' (non-standard) noise\cite{Tsirelson} that decides what the 
particles do when they meet. Constructing the transition probabilities 
of the generalized flow for the sticky boundary conditions in more than
one dimension is still an open problem.

\subsection{Zero-mode mechanism for scalar intermittency}

The statistics of fully turbulent velocities exhibits 
the phenomenon of intermittency, characterized by frequent appearance 
of strong signals interwoven with weak ones both in time and in space 
domain \cite{Frisch}. Similar, but even more pronounced effects 
characterize the statistics of advected scalars \cite{AHGA,MWAT}.
One of the motivations behind studying scalar advection in Kraichnan 
velocities, whose Gaussian statistics does not exhibit any 
intermittency, was a search for a mechanism of this enhanced scalar 
intermittency \cite{Kr94}. We shall briefly describe such a mechanism 
discovered subsequently \cite{ChFKL,GawKup,SS0} and its interpretation 
in terms of the related flow of Lagrangian particles.

\subsubsection{Passive scalar in Kraichnan velocities} 

For the scalar advected by smooth velocities, 
$$
  \theta(t,\bm R(t;t_0,\bm r_0))
\,=\,
  \theta(t_0,\bm r_0)
\,=\,
  \const
$$
In the Kraichnan velocity ensemble this means that the scalar field
has to satisfy the It\^o stochastic PDE
$$
  \dd_t \theta(t,\bm r)
\,=\,
 -\big(\bm v(t,\bm r)\dd t\big)\cdot \bm\nabla \theta(t,\bm r) 
 \,+\,\half D_0\,\bm\nabla^2 \theta(t,\bm r) \,\dd t\,.
$$
Indeed, the It\^o rule gives then
\begin{align*}
  \dd_t \theta(t,\bm R(t))
\,=&
 -\big(\bm v(t,\bm R(t))\dd t\big)\cdot \bm\nabla \theta(t,\bm R(t))
 \,+\,\half D_0 \bm\nabla^2 \theta(t,\bm R(t)) \,\dd t
\\
&+\,\big(\bm v(t,\bm R(t))\dd t\big)\cdot \bm\nabla \theta(t,\bm R(t))
 \,+\,\half D_0 \bm\nabla^2 \theta(t,\bm R(t)) \,\dd t
\\
&-D_0 \bm\nabla^2 \theta(t,\bm R(t)) \,\dd t
\,=\,0
\end{align*}
(the last term in the intermediate equality came from the mixed increments 
of $\theta(t)$ and of $\bm R(t)$). \,The corresponding Stratonovich 
stochastic PDE would be
$$
  \dd_t \theta(t,\bm r)
\,=\,
 -\big(\bm v(t,\bm r)\dd t\big)\cdot\bm\nabla \theta(t,\bm r)
$$
(without the diffusion term).
\,Similarly, for the scalar with a source, one obtains the It\^o equation
\qq
  \dd_t \theta(t,\bm r)
\,=\,
 -\big(\bm v(t,\bm r)\dd t\big)\cdot \bm\nabla \theta(t,\bm r) 
 \,+\,\half D_0\,\bm\nabla^2 \theta(t,\bm r) \,\dd t
 \,+\,g(t,\bm r)\,\dd t
\label{scs}
\qqq
Let us suppose that $\,g(t,\bm r)\,$ is also a Gaussian process,
independent from velocities, with mean zero and covariance
$$\big\langle g(t,\bm r)\,g(t',\bm r')\big\rangle\,=\,
\delta(t-t')\,\chi(\bm r-\bm r')\,.$$
For the product $\,\prod_{n=1}^N \theta(t,\bm r_n)$, \,one obtains
then from Eq.\,(\ref{scs}) the It\^o SDE
\qq
\dd_t \prod_{n=1}^N \theta(t,\bm r_n)
&=&
-\sum_{n=1}^N
    \big(\bm v(t,\bm r_n)\dd t\big)\cdot \bm\nabla \theta(t,\bm r_n)
    \prod_{n' \neq n} \theta(t,\bm r_{n'})
\,+\,\sum_{n=1}^N
    \half D_0 \bm\nabla^2 \theta(t,\bm r_n)
    \prod_{n' \neq n} \theta(t,\bm r_{n'})\,\dd t
\cr
&+&\sum_{m<n}
    D^{ij}(\bm r_m - \bm r_n)\,
      \nabla_i \theta(t,\bm r_m) \nabla_j \theta(t,\bm r_n) 
    \prod_{n' \neq n,m} \theta(t,\bm r_{n'})\,\dd t\cr
&+&\sum_{n=1}^N\big(g(t,\bm r_n)\dd t\big)\prod_{n' \neq n} 
\theta(t,\bm r_{n'})\,+\,\sum_{n<n'}\chi(\bm r_n-\bm r'_{n'})
\prod_{n'' \neq n,n'}\theta(t,\bm r_{n'})\,\dd t\,.
\nonumber
\qqq
For the expectation values, this gives the relation
$$
  \frac{\dd}{\dd t} \Big\langle \prod_{n=1}^N \theta(t,\bm r_n)\Big\rangle
\ =\ 
  \mathcal M_{\hspace{-0.03cm}_N}
\,\Big\langle\prod_{n=1}^N \theta(t,\bm r_n)\Big\rangle
 \ +\ \sum_{n<n'}
    \chi(\bm r_n-\bm r'_{n'})\,
\Big\langle \prod_{n'' \neq n,n'} \theta(t,\bm r_{n'})\Big\rangle\,,
$$
where the operator $\,\mathcal M_{\hspace{-0.03cm}_N}\,$ is the 
same as in Eq.\,(\ref{MN}) for the evolution of expectations 
of functions of positions of $\,N\,$ Lagrangian particles.

\subsubsection{Zero modes and anomalous scaling}

Under steady forcing, the scalar statistics reaches a stationary state
in the weakly compressible phase with direct scalar energy cascade. 
In this state,
$$
  \mathcal M_{\hspace{-0.03cm}_N}
    \,\Big\langle \prod_{n=1}^N \theta(t,\bm r_n)\Big\rangle\ =\ 
 -\sum_{n,n'} \chi(\bm r_n-\bm r_{n'}))\,
    \Big\langle \prod_{n'' \neq n,n'} \theta(t,\bm r_{n'})\Big\rangle\,.
$$
The above equations determine the scalar correlations 
$\,\big\langle \prod \theta(t,\bm r_n)\big\rangle\,$
inductively in $\,N$, \,up to zero modes of the operators 
$\,\mathcal M_{\hspace{-0.03cm}_N}$.
\,It appears that if the source $\,g\,$ is concentrated mostly 
at long distances then the short distance scaling of the scalar 
``structure
  functions'' $\,\,S_{\hspace{-0.03cm}_N}(\bm r)
= \big\langle (\theta(t,\bm r) - 
\theta(t,\bm 0))^N\big\rangle$,
\qq
  S_{\hspace{-0.03cm}_N}(\bm r)
\ \propto\ |\bm r|^{\zeta_N}\ \ \quad\text{for small \,$|\bm r|$}\,,
\qqq
is determined by the contribution of the scaling zero-modes 
$\,f_{\hspace{-0.03cm}_N}\,$ of
$\,\mathcal M_{\hspace{-0.03cm}_N}\m$:
$$
  \mathcal M_{\hspace{-0.03cm}_N} f_{\hspace{-0.03cm}_N}
\,=\,
  0
\,,\ \ \quad
  f_{\hspace{-0.03cm}_N}(\lambda \underline{\bm R})
\,=\,
  \lambda^{\zeta_{\hspace{-0.03cm}_N}} f_{\hspace{-0.03cm}_N}(\underline{\bm R})\,.
$$
Recall that the functions of the $N$ Lagrangian trajectories evolve by
Eq.\,(\ref{MN}).
\,The zero-modes of $\mathcal M_{\hspace{-0.03cm}_N}$ correspond 
to functions $\,f(\underline{\bm R}(t))\,$ of the $N$-particle process
that are conserved in mean:
$$
  \big\langle
    f(\underline{\bm R}(t;\underline{\bm r}_0))
  \big\rangle\,=\,  f(\underline{\bm r}_0)\,.
$$
They are martingales of the effective multi-particle diffusions.
\vskip 0.1cm

The scaling zero-modes of $\,\mathcal M_{\hspace{-0.03cm}_N}\,$ 
giving the leading contributions to the structure function 
$\,S_{\hspace{-0.03cm}_N}(\bm r)\,$ have been found for homogeneous and
isotropic flows 
in the leading order of the perturbation expansion in powers of $\,\xi\,$
\cite{GawKup,BGK} and of $\,\frac{1}{d}\,$ \cite{ChFKL,ChFalk} in
the years 1995-6.  The result was:
\qq
  \zeta_N
=
  \begin{cases}
    \,\,\frac{N}{2}(2-\xi) - \frac{N(N-2)(1+2\wp)\,\xi}{2(d+2)} + O(\xi^2)
  \\[1ex]
    \,\underbrace{\,\textstyle
      \frac{N}{2}(2-\xi)}_{{\rm dimensional}\atop{\rm scaling}}
    
   -\underbrace{\textstyle
      \frac{N(N-2)(1+2\wp)\,\xi}{2d} + O(\frac{1}{d^2})}_\text{anomaly}
  \\[-6ex]
  \end{cases}
\label{pert}
\qqq
\vskip 0.65cm

\noindent Terms to the \third order are known \cite{StPeter}.
The fact that $\,\zeta_{\hspace{-0.03cm}_N} < \frac{N}{2}\zeta_2\,$ means 
that $$\,\frac{S_{\hspace{-0.03cm}_N}(\bm r)}{S_2(\bm r)^{N/2}}\,$$ 
becomes large for small $\,|\bm r|$.  \,This
signals the presence of long tails in the stationary distribution of
the scalar increments $\,\big(\theta(t,\bm r) - \theta(t,\bm 0)\big)$, 
\,getting more and more pronounced at short distances.  This is a signal 
of small scale intermittency of the scalar statistics observed 
also in other turbulent states.
\,The perturbative results (\ref{pert}) were confirmed by 
numerical simulations, see the figure from \cite{FMNV}

\begin{figure}[h!]
\vspace{0.1cm}
\begin{center}
\mbox{\hspace{0.0cm}\epsfig{file=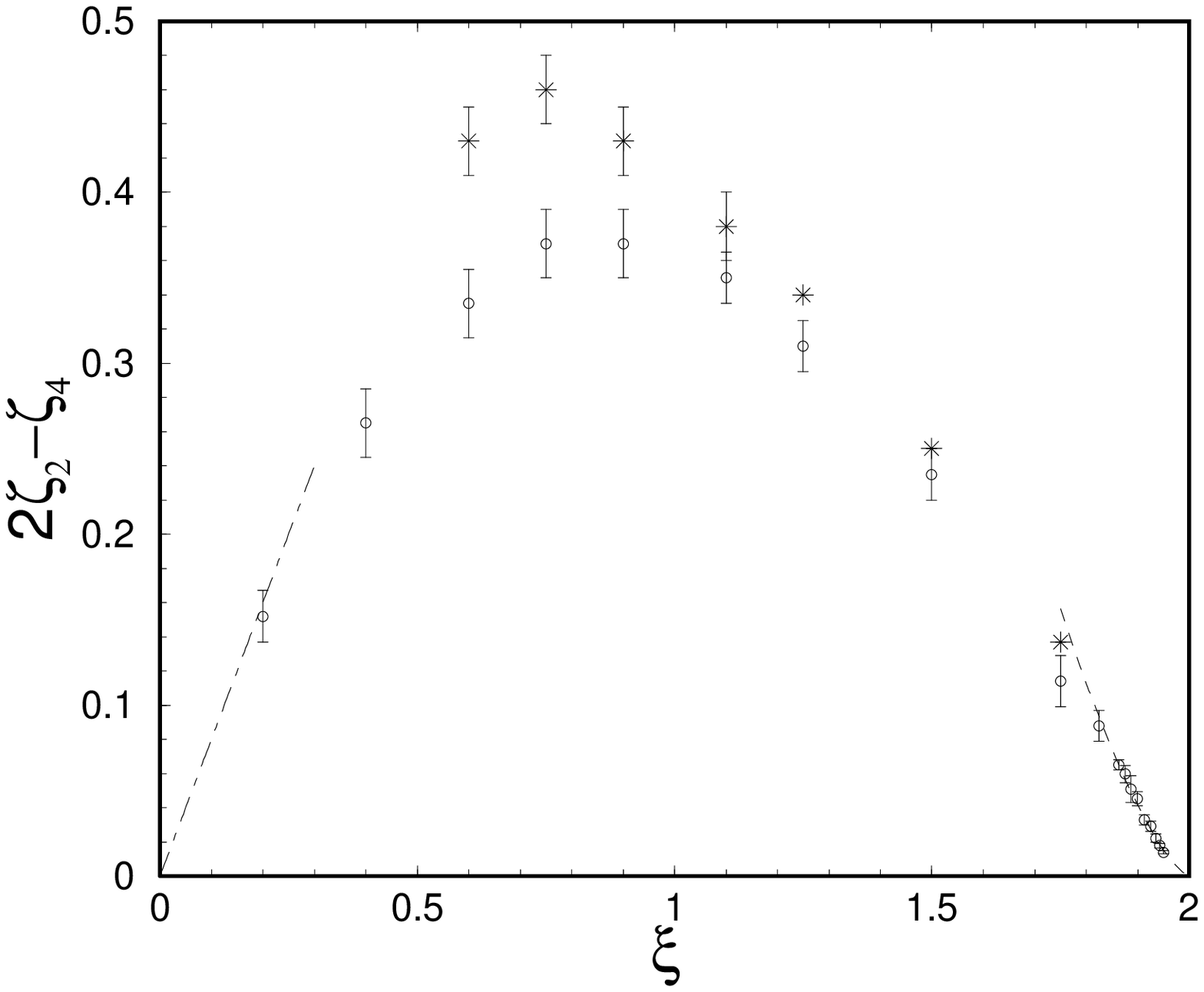,height=4.6cm,width=5.3cm}}
\end{center}
\vspace{-0.2cm}
\end{figure}

\noindent representing the results for $2\zeta_2-\zeta_4$ as a function 
of $\xi$ in two- (upper points) and three-dimensional (lower points) 
incompressible Kraichnan model, with the broken line in the lower left 
corner representing the perturbative result (\ref{pert}) for $d=3$.
Zero-modes were also shown numerically to be also responsible for the
intermittency of scalar transport by the inverse cascade of 2D
turbulence \cite{CelVerg}.

\begin{conc}
In the non-smooth Kraichnan velocities modeling the inertial range
turbulence, Lagrangian particles exhibit non-standard behaviors 
impossible in differentiable dynamical systems but expected 
to be common in non-differentiable ones with H\"older-continuous
vector fields. These behaviors include explosive separation 
of trajectories generating their spontaneous randomness 
and an implosive collapse of deterministic trajectories, 
with possible stickiness. Such unusual behaviors of trajectories 
are source of non-equilibrium transport phenomena involving cascades 
with non-zero fluxes of conserved quantities. They condition the 
appearance of short-scale intermittency in the direct cascade of scalars 
advected by fully turbulent flows. The Kraichnan model allowed 
to associate the scalar intermittency to hidden statistical conservation 
laws of multi-particle evolution: the ``zero modes''. The zero-mode scenario 
permitted perturbative calculations of the anomalous scaling exponents 
of the scalar structure functions in this model. Similar mechanism 
is believed to be responsible for intermittency is other problems with 
dynamics governed by linear evolution equations \cite{ABCPV}.
\end{conc}

\newpage

\nsection{Final remarks}

The passive transport in turbulent flows is related to the behavior of
Lagrangian or inertial particles carried by the fluid.  In the
regime where velocities are smooth in space, i.e.\ for intermediate 
Reynolds numbers, the particle dynamics provides examples of random 
differential dynamical systems and may be studied with tools of 
the chaotic dynamical systems theory. It requires, nevertheless, subtle 
information about rare dynamical events (multiplicative large deviations). 
The statistics of rare fluctuations obeys fluctuation relations, sharing 
this property with other non-equilibrium systems \cite{CheGaw}.
\vskip 0.1cm

In the regime of large Reynolds numbers, i.e.\ in the inertial range,
the motion of particles may be viewed as providing an example of random
non-differentiable dynamical system.  As the work on the Kraichnan
model has shown, such systems lead to unconventional flows with
explosive trajectory separation or implosive trajectory trapping.
The exotic behavior of particle trajectories has dramatic effect on 
the transport properties and conditions the appearance of cascades 
of conserved quantities and of intermittency related to hidden 
conservation laws. 
\vskip 0.1cm

Some of the lessons from studying passive turbulent transport 
remain valid for reactive particles like water droplets or chemical
or biological agents \cite{FFS,TMGK}. How many of them may be
be transformed to insights about turbulence itself remains to be 
seen \cite{ABBPT}.


\begin{thebibliography}{bib}
\bibitem{StPeter}
L. Ts. Adzhemyan, N. V. Antonov, V. A. Barinov, Yu. S. Kabrits, A. N. Vasil'ev,
``Calculation of the anomalous exponents in the rapid-change model 
of passive scalar advection to order $\varepsilon^{3}$'',
Phys. Rev. {\bf E 64} (2001), 056306/1-28
\bibitem{ABBPT}
L. Angheluta, R. Benzi, L. Biferale, I. Procaccia, F. Toschi,
``Anomalous scaling exponents in nonlinear models of turbulence'',
 Phys. Rev. Lett. {\bf 97} (2006), 160601/1-4 
\bibitem{AHGA}
R. A. Antonia, E. Hopfinger, Y. Gagne, F. Anselmet,
``Temperature structure functions in turbulent shear flows'', 
Phys. Rev. {\bf A 30} (1984), 2704-2707
\bibitem{ABCPV}
I. Arad, L. Biferale, A. Celani, I. Procaccia and M. Vergassola,
``Statistical conservation laws in turbulent transport'',
Phys. Rev. Lett. {\bf 87} (2001),  164502/1-4
\bibitem{LArnold}
L. Arnold, {\em Random Dynamical Systems}, Springer, Berlin 2003
\bibitem{BalkFalkFoux}
E. Balkovsky, G. Falkovich, A. Fouxon, 
``Clustering of inertial particles in turbulent flows'',
chao-dyn/9912027 and ``Intermittent distribution
of inertial particles in turbulent flows'', Phys. Rev. Lett. {\bf 86} (2001),
2790-2793
\bibitem{BalkFoux}
E. Balkovsky, A. Fouxon,
``Universal long-time properties of Lagrangian statistics 
in the Batchelor regime and their application to the passive 
scalar problem'', 
Phys.Rev. E, {\bf 60}, (1999) 4164-4174
\bibitem{BalkFouxLeb}
E. Balkovsky and A. Fouxon and V. Lebedev,
``On the Turbulent Dynamics of Polymer Solutions'',
Phys. Rev. Lett. {\bf 84} (2000), 4765-4768
\bibitem{BCG}
M. M. Bandi, J. R. Cressman Jr., W. I. Goldburg,
``Test of the Fluctuation Relation in compressible turbulence on 
a free surface'', J. Stat. Phys. {\bf 130} (2008), 27-38
\bibitem{Batch}
G. K. Batchelor, {\em An Introduction to Fluid Dynamics}, Cambridge 
University Press 1967
\bibitem{BaxStr}
P. H. Baxendale, D. W. Stroock,
``Large deviations and stochastic flows of diffeomorphisms'', 
Prob. Theor.\& Rel. Fields {\bf 80} (1988), 169-215
\bibitem{Bec}
J. Bec, 
``Multifractal concentrations of inertial particles in smooth random flows'',
J. Fluid Mech. {\bf 528} (2005), 255-277
\bibitem{BBCLMT}
J. Bec, L. Biferale, M. Cencini, A. Lanotte, S. Musacchio, F. Toschi,
``Heavy particle concentration in turbulence at dissipative and 
inertial scales'',
Phys. Rev. Lett. {\bf 98} (2007), 084502/1-4
\bibitem{BCH0}
J. Bec, M. Cencini, R. Hillerbrand, 
``Heavy particles in incompressible flows: the large Stokes number 
asymptotics'', Physica {\bf D 226} (2007), 11-22
\bibitem{BCH}
J. Bec, M. Cencini, R. Hillerbrand,
``Clustering of heavy particles in random self-similar flow''
Phys. Rev. {\bf E 75} (2007), 025301(R)/1-4
\bibitem{BGH}
J. Bec, K. Gaw\c{e}dzki, P. Horvai, 
``Multifractal clustering in compressible flows'', 
Phys. Rev. Lett. {\bf 92} (2004), 224501-2240504
\bibitem{BLTP}
R. Benzi, B. Levant, I. Procaccia, E. S. Titi,
``Statistical properties of nonlinear shell models of turbulence from 
linear advection models: rigorous results'',
arXiv:nlin.CD/0612033
\bibitem{BGK}
D. Bernard, K. Gaw\c{e}dzki, A. Kupiainen,
``Anomalous scaling in the N-point functions of a passive scalar'',
Phys. Rev. {\bf E 54} (1996), 2564-2572
\bibitem{slow}
D. Bernard, K. Gaw\c{e}dzki, A. Kupiainen,
``Slow modes in passive advection'',
J. Stat. Phys. {\bf 90} (1998), 519-569
\bibitem{BHAC}
R. B. Bird, C. F. Curtiss, R. C. Armstrong, O. Hassager,
{\em Dynamics of Polymeric Liquids, Vol. 2, Kinetic Theory}, 
Wiley, New York 1987
\bibitem{BDES}
G. Boffetta, J. Davoudi, B. Eckhardt, J. Schumacher,
``Lagrangian tracers on a surface flow: the role of time correlations'',
Phys. Rev. Lett. {\bf 93} (2004), 134501/1-4
\bibitem{BJPV}
T. Bohr, M. H. Jensen, G. Paladin, A. Vulpiani,
{\em Dynamical Systems Approach to Turbulence},
Cambridge University Press 1998
\bibitem{BDD}
G. Boffetta, J. Davoudi, F. De Lillo,
``Multifractal clustering of passive tracers on a surface flow'',
Europhys. Lett., {\bf 74} (2006), 62-68 
\bibitem{BonGalGent}
F. Bonetto, G. Gallavotti, G. Gentile,
``A fluctuation theorem in a random environment'',
mp\_arc/06-139
\bibitem{CelVerg}
A. Celani, M. Vergassola,
``Statistical Geometry in Scalar Turbulence'',
Phys. Rev. Lett. {\bf 86} (2001), 424-427 
\bibitem{Chertkov}
M. Chertkov,
``Polymer Stretching by Turbulence'',
Phys. Rev. Lett. {\bf 84} (2000), 4761-4764
\bibitem{ChFalk}
M. Chertkov, G. Falkovich,
``Anomalous scaling exponents of a white-advected passive scalar'', 
Phys. Rev. Lett. {\bf 76} (1996), 2706-2709
\bibitem{ChFKL}
M. Chertkov, G. Falkovich, I. Kolokolov, V. Lebedev,
``Normal and anomalous scaling of the fourth-order
correlation function of a randomly advected scalar'', 
Phys. Rev. {\bf E 52} (1995), 4924-4941
\bibitem{ChKolVerg}
M. Chertkov, I. Kolokolov, M. Vergassola,
``Inverse versus direct cascades in turbulent advection'',
Phys. Rev. Lett. {\bf 80} (1998), 512-515
\bibitem{CheGaw}
R. Chetrite, K. Gaw\c{e}dzki, ``Fluctuation
relations for diffusion processes'', Commun. Math. Phys., in print,
available online
\bibitem{CDG}
R. Chetrite, J.-Y. Delannoy, K. Gaw\c{e}dzki, 
``Kraichnan flow in a square: an example of integrable chaos'',
arXiv:nlin.CD/0606015, J. Stat. Phys. in press
\bibitem{ChevMan}
L. Chevillard, C. Meneveau, ``Intermittency and universality in
a Lagrangian model of velocity gradients in three-dimensional turbulence'',
C. R. Mechanique {\bf 335} (2007), 187-193 
\bibitem{EV1}
W. E, E. Vanden Eijnden,
``Generalized flows, intrinsic stochasticity, and turbulent transport'',
Proc. Nat. Acad. Sci. {\bf 97} (2000), 8200-8205
\bibitem{EV2}
W. E, E. Vanden Eijnden,
``Turbulent Prandtl number effect on passive scalar advection'', 
Physica D {\bf 152-153} (2001), 636-645
\bibitem{Eck}
J.-P. Eckmann,
``Roads to turbulence in dissipative dynamical systems''
Rev. Mod. Phys. {\bf 53} (1981), 643 - 654
\bibitem{EvSear1}
D. J. Evans, D. J. Searles,
``Equilibrium microstates which generate the second law violating
steady states''
Phys. Rev. {\bf E 50} (1994), 1645-1648
\bibitem{EvSear2}
Evans, D. J., Searles, D. J.:
{\it The fluctuation theorem}. Adv. in Phys., {\bf 51} (2002), 1529-1585
\bibitem{FFS}
G. Falkovich, A. Fouxon, M. G. Stepanov, ``Acceleration of rain 
initiation by cloud turbulence'', Nature {\bf 419}  (2002), 151-154
\bibitem{FGV}
G. Falkovich, K. Gawedzki, M. Vergassola,
``Particles and fields in fluid turbulence'',
Rev. Mod. Phys. {\bf 73} (2001), 913-975
\bibitem{FouxHorv}
I. Fouxon, P. Horvai, ``Fluctuation relation and pairing rule 
for Lyapunov exponents of inertial particles in turbulence'',
J. Stat. Mech.: Theor. Experim. {\bf 8} (2007), L08002/1-9
\bibitem{FH}
I. Fouxon and P. Horvai, ``Separation of heavy particles in turbulence'',
Phys. Rev. Lett. {\bf 100} (2008), 040601/1-4
\bibitem{Frisch}
U. Frisch: {\em Turbulence: the Legacy of A. N. Kolmogorov},
Cambridge University Press, Cambridge 1995
\bibitem{FMNV}
U. Frisch, A. Mazzino, A. Noullez, M. Vergassola, 
``Lagrangian method for multiple correlations in passive scalar advection'', 
Phys. Fluids {\bf 11} (1999), 2178-2186
\bibitem{GalCoh}
G. Gallavotti, E. D. G. Cohen,
``Dynamical ensembles in non-equilibrium statistical mechanics'',
Phys. Rev. Lett. {\bf 74} (1995), 2694-2697
\bibitem{Warwick}
K. Gaw\c{e}dzki, ``Soluble models of turbulent transport'',
in: Non-equilibrium Statistical Mechanics and Turbulence, Eds. S. Nazarenko, 
O. V. Zaboronski, Cambridge University Press 2008
\bibitem{GawHorv}
K. Gaw\c{e}dzki, P. Horvai,
``Sticky behavior of fluid particles in the compressible Kraichnan model'',
J. Stat. Phys. {\bf 116} (2004), 1247-1300
\bibitem{GawKup}
K. Gaw\c{e}dzki, A. Kupiainen, 
``Anomalous scaling of the passive scalar'', 
Phys. Rev. Lett. {\bf 75} (1995), 3834-3837
\bibitem{GV}
K. Gaw\c{e}dzki, M. Vergassola,
``Phase transition in the passive scalar advection'', 
Physica {\bf D 138} (2000), 63-90 
\bibitem{GrRh}  
I. S. Gradshteyn, I. M. Ryzhik, Table of Integrals, 
Series and Products. $7^{\mathrm{th}}$ Edition. Eds. Jeffrey, 
A., Zwillinger, D., Academic Press, New York 2007 
\bibitem{Grass}
P. Grassberger, R. Baddi, A. Politi,
``Scaling laws for invariant measures on hyperbolic and nonhyperbolic 
attractors'', 
J. Stat. Phys. {\bf 51} (1988), 135-178
\bibitem{Halp}
B. I. Halperin, ``Green's functions for a particle
in a one-dimensional random potential'', Phys. Rev. {\bf 139} (1965), 
A104-A117
\bibitem{Horv}
P. Horvai,
``Lyapunov exponent for inertial particles in the 2D Kraichnan model 
as a problem of Anderson localization with complex valued potential'',
arXiv:nlin.CD/0511023
\bibitem{Jarz1}
Jarzynski, C.: {\it A nonequilibrium equality for free energy differences}. 
Phys. Rev. Lett. {\bf 78} (1997), 2690-2693 
\bibitem{K41}
A. N. Kolmogorov, ``The local structure of
turbulence in incompressible viscous fluid for very large
Reynolds' numbers'', C. R. Acad. Sci. URSS {\bf 30} (1941), 301-305
\bibitem{Kr68}
R. H. Kraichnan, 
``Small-scale structure of a scalar field convected by turbulence'', 
Phys. Fluids {\bf 11} (1968), 945-963
\bibitem{Kr94}
 R. H. Kraichnan, ``Anomalous scaling of a randomly advected
passive scalar'', Phys. Rev. Lett. {\bf 72} (1994), 1016-1019
\bibitem{LeJRai1}
Y. Le Jan, O. Raimond, 
``Integration of Brownian vector fields'',
Ann. Probab. {\bf 30} (2002), 826-873
\bibitem{LeJRai2}
Y. Le Jan, O. Raimond, 
``Flows, coalescence and noise'',
Ann. Probab. {\bf 32} (2004), 1247-1315
\bibitem{LGP}
I. M. Lifshitz, S. Gredeskul, L. Pastur, {\em Introduction
to the Theory of Disordered Systems}, Wiley N.Y. 1988
\bibitem{MK}
A. J. Majda and P. R. Kramer,
``Simplified models for turbulent diffusion:  
Theory, numerical modelling and physical phenomena'', 
Physics Reports {\bf 314} (1999), 237-574
\bibitem{Max83}
M. Maxey, J. Riley, ``Equation of motion of a small 
rigid sphere in a nonuniform flow'',
Phys. Fluids {\bf 26} (1983), 883-889
\bibitem{MWDWL}
B. Mehlig, M. Wilkinson, K. Duncan, T. Weber, M. Ljunggren,
``On the aggregation of inertial particles in random flows''
Phys. Rev. {\bf E 72} (2005), 051104/1-10
\bibitem{MWAT}
F. Moisy, H. Willaime, J. S. Andersen, P. Tabeling,
``Passive Scalar Intermittency in Low Temperature Helium Flows'',
Phys. Rev. Lett. {\bf 86} (2001), 4827 - 4830
\bibitem{Oksen}
B. Oksendal,
{\em Stochastic Differential Equations}, 6th ed., Universitext, 
Springer, Berlin 2003
\bibitem{OlshPer}
M. A. Olshanetsky, A. M. Perelomov:
``Quantum integrable systems related to Lie algebras''
Phys. Rep. {\bf 94} (1983), 313-404
\bibitem{Osel}
V. I. Oseledec, ``Multiplicative ergodic theorem: characteristic
Lyapunov exponents of dynamical systems'',
Trudy Moskov. Mat. Ob$\check{\rm s}\check{\rm c}$. {\bf 19}
(1968), 179-210
\bibitem{Piterb}
L. Piterbarg, ``The top Lyapunov exponent for a stochastic flow modelling
the upper ocean turbulence'', SIAM J. Appl. Math. {\bf 62} (2001), 777-800
\bibitem{Richar}
L. F. Richardson,
``Atmospheric diffusion shown on a distance-neighbour graph'',
Proc. R. Soc. Lond. {\bf A 110} (1926), 709-737
\bibitem{Risk}
H. Risken, 
{\em The Fokker Planck equation}, Springer, Berlin, 1989
\bibitem{Ruelle1}
D. Ruelle, 
``Ergodic theory of differentiable dynamical systems'',
Publications Math\'ematiques de l'IH\'ES {\bf 50} (1979), 275-320
\bibitem{Ruelle2}
D. Ruelle, 
``Positivity of entropy production in nonequilibrium statistical 
mechanics'',
J. Stat. Phys. {\bf 85} (1996), 1-23
\bibitem{Ruelle3}
D. Ruelle, 
``Positivity of entropy production in the presence of a random 
thermostat'',
J. Stat. Phys. {\bf 86} (1997), 935-951 
\bibitem{Saw}
B. L. Sawford, ``Turbulent relative dispersion'', Annual Rev. Fluid Mech.
{\bf 33} (2001), 289-317
\bibitem{SS0}
B. I. Shraiman, E. D. Siggia, 
``Anomalous scaling of a passive scalar in turbulent flow'', 
C.R. Acad. Sci.{\bf 321} (1995), 279-284
\bibitem{SS}
B. I. Shraiman, E. D. Siggia, ``Scalar turbulence'', Nature {\bf 405} (2000),
639-646
\bibitem{TMGK}
T. T\'el, A. de Moura, C. Grebogi, G. K\'arolyi,
``Chemical and biological activity in open flows: A dynamical 
system approach'',
Phys. Rep. {\bf 413} (2005), 91-196
\bibitem{Tsirelson}
B. Tsirelson,
``Nonclassical stochastic flows and continuous products'',
Probab. Surveys {\bf 1} (2004), 173-298
\bibitem{Warh}
Z. Warhaft, ``Passive scalars in turbulent flows'', Annual Rev. Fluid Mech.
{\bf 32} (2000), 203-240
\bibitem{WiggOtt}
S. Wiggins, J. M. Ottino, ``Foundations of chaotic mixing'', Phil. Trans.
R. Soc. Lond. A {\bf 362} (2004), 937-970 
\bibitem{WM}
M. Wilkinson, B. Mehlig, ``Path coalescence transition and its applications'',
Phys. Rev. {\bf E 68} (2003), 040101/1-4 
\end{thebibliography}
\end{document}